\documentclass[11pt]{article}
\usepackage{geometry}[top=1in,left=1in,right=1in,bottom=1in]
\usepackage[english]{babel}
\usepackage[utf8]{inputenc}
\usepackage[OT4]{fontenc}
\usepackage{amsmath}
\usepackage{multirow}
\usepackage{setspace}
\usepackage{amssymb}
\usepackage[noline,boxed]{algorithm2e}
\usepackage[skins]{tcolorbox}
\usepackage{graphicx}
\graphicspath{{Figures/}}
\usepackage{subcaption}
\usepackage{tikz}
\usepackage{geometry}
\usepackage{url}
\usepackage{mathtools}
\usepackage{mleftright}
\usepackage{IEEEtrantools}

\usepackage{marginnote}
\newcommand\preceqdot{\mathrel{\ooalign{$\preceq$\cr
\hidewidth\raise0.225ex\hbox{$\cdot\mkern0.5mu$}\cr}}}
\usepackage{graphicx, tikz-cd}
\usepackage{enumerate}
\usepackage{amssymb,  latexsym}
\usetikzlibrary{shapes, arrows, intersections}

\newcommand{\eps}{\varepsilon}
\newcommand{\cC}{\mathcal{C}}
\newcommand{\cP}{\mathcal{P}}
\newcommand{\Cdec}{\cC_{\mathrm{dec}}}
\newcommand{\PB}{P_{\mathrm{B}}}
\newcommand{\Pamb}{P_{\mathrm{amb}}}
\newcommand{\Pue}{P_{\mathrm{ue}}}
\newcommand{\ymin}{y_{\mathrm{min}}}
\newcommand{\ymax}{y_{\mathrm{max}}}

\newcommand{\Xhat}{\widehat{X}}

\newcommand{\Bh}{Z}

\newcommand{\kEC}{\mathop{\mbox{$k$-$\mathrm{EC}$}}}
\newcommand{\kSC}{\mathop{\mbox{$k$-$\mathrm{SC}$}}}
\DeclareMathOperator{\wt}{wt}
\DeclareMathOperator*{\EE}{\mathbb{E}}

\DeclareMathOperator{\Ber}{Ber}

\DeclareMathOperator*{\argmax}{arg\,max}
\DeclareMathOperator{\supp}{supp}
\DeclareMathOperator{\sgn}{sgn}
\DeclareMathOperator{\LHS}{LHS}

\usepackage{hyperref}
\usepackage{amsthm}
\usepackage{cleveref}

\newtheorem{theorem}{Theorem}
\newtheorem{example}[theorem]{Example}
\newtheorem{proposition}[theorem]{Proposition}
\newtheorem{corollary}[theorem]{Corollary}
\newtheorem{lemma}[theorem]{Lemma}
\newtheorem{claim}[theorem]{Claim}

\theoremstyle{remark}
\newtheorem{remark}[theorem]{Remark}

\theoremstyle{definition}
\newtheorem{definition}[theorem]{Definition}

\title{Generalized Samorodnitsky noisy function inequalities,\\
with applications to error-correcting codes}

\author{
Olakunle Abawonse\footnote{AIMS Rwanda. Email:\url{olakunle.abawonse@aims.ac.rw}.}
\and
Jan Hązła\footnote{AIMS Rwanda. Email:\url{jan.hazla@aims.ac.rw}.
O.A.~and J.H.~were
supported by the Alexander von Humboldt Foundation German research chair
funding and the associated DAAD projects No. 57610033
and 57761435.}
\and
Ryan O'Donnell \footnote{Carnegie Mellon University. Email:\url{odonnell@cs.cmu.edu}.}
}

\date{}

\begin{document}

\maketitle

\begin{abstract}
An inequality by Samorodnitsky states that if $f : \mathbb{F}_2^n \to \mathbb{R}$ is a nonnegative boolean function, and $S \subseteq [n]$ is chosen by randomly including each coordinate with probability a certain $\lambda = \lambda(q,\rho) < 1$, then
\begin{equation}
    \log \|T_\rho f\|_q \leq \EE_{S} \log \|\EE(f|S)\|_q\;.
\end{equation}
Samorodnitsky's inequality has several applications to the theory of error-correcting codes. Perhaps most notably, it can be used to show that \emph{any} binary linear code (with minimum distance $\omega(\log n)$) that has vanishing decoding error probability on the BEC$(\lambda)$ (binary erasure channel) also has vanishing decoding error on \emph{all} memoryless symmetric channels with capacity above some $C = C(\lambda)$.

Samorodnitsky determined the optimal $\lambda = \lambda(q,\rho)$ for his inequality in the case that $q \geq 2$ is an integer.  In this work, we generalize the inequality to $f : \Omega^n \to \mathbb{R}$ under any product probability distribution $\mu^{\otimes n}$ on $\Omega^n$; moreover, we determine the optimal value of $\lambda = \lambda(q,\mu,\rho)$ for any real $q \in [2,\infty]$, $\rho \in [0,1]$, and distribution~$\mu$.
As one consequence, we obtain the aforementioned coding theory result for linear codes over \emph{any} finite alphabet.
\end{abstract}

\section{Introduction}

In 2019, Samorodnitsky~\cite{Sam19} proved a remarkable inequality of the following form:
\newtheorem*{samtheorem}{Samorodnitsky Inequality}
\begin{samtheorem}
    \emph{For all real $q > 1$  and $0 < \rho < 1$, there exists a certain $\lambda = \lambda(q,\rho) < 1$ such that the following holds for all nonnegative $f : \mathbb{F}_2^n \to \mathbb{R}$:
    \begin{equation}
        \log\|T_\rho f\|_q\le \EE_{S\sim\lambda} \log\|\EE(f|S)\|_q.
    \end{equation}}
\end{samtheorem}

Here $T_\rho$
is the noise operator with correlation $\rho\in(0,1)$, 
$\EE(f|S)$ for $S\subseteq\{1,\ldots,n\}$ denotes the conditional expectation operator applied to $f$, and $S\sim\lambda$ indicates that each element
is included in the random set $S$ independently with probability $\lambda$.
(See~\Cref{sec:induction} for full definitions.)

Samorodnitsky showed that his inequality holds for general $n$ if and only if it holds for $n = 1$, in which case the inequality is equivalent to
\begin{equation} \label{eqn:f2-n1}
    \|T_\rho f\|_q \leq \|f\|_1^{1-\lambda} \|f\|_q^{\lambda}\;.
\end{equation}
(See below for a comparison with the celebrated hypercontractive inequality~\cite{Bon70}.)
Moreover, in a subsequent work~\cite{Sam20}, Samorodnitsky  determined the optimal value of $\lambda$ whenever $q \geq 2$ is an integer.  Specifically, for integer $q$, he showed the worst case is achieved by the ``dictator'' function $f : \mathbb{F}_2 \to \{0,1\}$, $x \mapsto x$.

In this work, we give an optimal generalization of the Samorodnitsky inequality to all product probability spaces and all real $q \geq 2$.
\begin{theorem} \label{thm:main}
    (Generalized Samorodnitsky Inequality.) 
    Let $\mu$ be a (full-support) probability distribution on finite space $\Omega$, and let $\mu^*\coloneq \min_{x\in\Omega}\mu(x)$.
    Then for all $q \in [2,\infty]$ and $\rho \in (0,1)$, there exists a certain $\lambda = \lambda(q, \mu^*, \rho) \in (0,1)$ such that the following holds for all nonnegative $f : \Omega \to  \mathbb{R}$:
    \begin{equation} \label{eqn:generaln1}
        \|T_\rho f\|_q \leq \|f\|_1^{1-\lambda} \|f\|_q^{\lambda}
    \end{equation}
    (where $T_\rho$ and $\| \cdot \|_q$,$\|\cdot\|_1$ are defined 
    vis-a-vis~$\mu$).
    The inequality is tight if and only if $f : \Omega \to \mathbb{R}$ is constant or if it is (any nonnegative multiple of) the indicator of~$x^*$, where $\mu(x^*) = \mu^*$.
    Moreover, for any $n$, the inequality generalizes to  nonnegative $f : \Omega^n \to \mathbb{R}$ in the following form:
    \begin{equation}\label{eqn:generaln}
        \log\|T_\rho f\|_q\le \EE_{S\sim\lambda} \log\|\EE(f|S)\|_q\;.
    \end{equation}
\end{theorem}

\medskip

Knowing that the indicator function for~$x^*$ achieves the worst case for the inequality, one can  immediately write down the 
(slightly unwieldy) formula for the optimal $\lambda$:
\begin{equation}
\label{eqn:optimallam}
    \lambda(q,\mu^*,\rho) \coloneqq 
    \tfrac{1}{q-1} \log_k\Bigl(\tfrac{1}{k} (1+(k-1) \rho)^q + (1-\tfrac{1}{k})(1-\rho)^q\Bigr), \quad k \coloneqq \tfrac{1}{\mu^*} \in [2, \infty).
\end{equation}
Since it holds
$1-\lambda =
-\frac{1}{q-1}\log_k\left(\left(\rho+\frac{1-\rho}{k}\right)^q+(k-1)\left(\frac{1-\rho}{k}\right)^q\right)$, if $k$ is integer, then $1-\lambda$
is equal to 
the Rényi entropy of order $q$
with logarithm base $k$ for a distribution
with probabilities
$\left\{\rho+\frac{1-\rho}k,\frac{1-\rho}{k},\ldots,\frac{1-\rho}{k}\right\}$.

We also have the following simpler formulas in the case of $q = 2, \infty$:
\begin{align}     \lambda(2,\mu^*,\rho) &= \log_k(1+(k-1)\rho^2)\;,\\
\lambda(\infty,\mu^*,\rho) &= \log_k(1+(k-1)\rho)\;.
\end{align}
This yields the following alternative formulation of the $q = 2$ case:
\begin{corollary}   \label{cor:main2}
    Let $\mu$ be a (full-support) probability distribution on finite space $\Omega$, let $\mu^*\coloneq \min_{x\in\Omega}\mu(x)$, and write $k\coloneq \frac{1}{\mu^*}$.
    Then for all $\lambda \in (0,1)$, if we define
    \begin{equation}    \label{eqn:zk}
        z = z_k(\lambda) = \frac{k^\lambda - 1}{k-1} \in (0,1)
    \end{equation}
    (which has the asymptotics  $z_k(\lambda) = \frac{\ln k}{k-1} \lambda + O(\lambda^2)$ and $z_k(1-\delta) = 1 - (\frac{k}{k-1} \ln k)\delta + O(\delta^2)$),
    it holds for all nonnegative $f : \Omega^n \to  \mathbb{R}$ that
    \begin{equation}
            \mathrm{Stab}_z[f] \le \mathrm{GM}_{S\sim\lambda} \|\EE(f|S)\|_2^2\ ,
    \end{equation}
    where $\mathrm{Stab}_z[f]$ is the ``noise stability'' $\|T_{\sqrt{z}} f\|_2^2$, and $\mathrm{GM}$ denotes the geometric mean.
\end{corollary}

\subsection{Hypercontractivity and strong data processing inequalities}

In the form of \eqref{eqn:f2-n1}, Samorodnitsky's inequality bears a strong resemblance to Bonami's \emph{hypercontractive inequality}~\cite{Bon70}, which plays a major role in analysis of boolean functions. (See, e.g., \cite[Chapters 9,10]{OD14}, which includes major applications such as the KKL Theorem and the Friedgut--Bourgain sharp threshold theorems.)  In general, hypercontractive inequalities for real-valued functions $f$ on an $n$-fold product of probability spaces $(\Omega, \mu)$ read as 
\begin{equation}
    \|T_\rho f\|_q \leq \|f\|_p\;,
\end{equation}
with $1 < p < q$ and a certain value of $\rho = \rho(p,q,\mu)$.
These inequalities are also known to generically reduce to the case of $n = 1$ and nonnegative~$f$.
Like Samorodnitsky's inequality, they quantify in some way the ``smoothing'' effect that the noise operator $T_\rho$ has on functions.

For a more detailed comparison, we start by recalling that when the domain is the uniform distribution on $\mathbb{F}_2$, Bonami showed the optimal value for $\rho$ satisfies $\rho^2 = \frac{p-1}{q-1}$.  As far as we know, this result is incomparable with Samorodnitsky's inequality.  It can be used to get a weaker form, though: from hypercontractivity and Riesz--Thorin interpolation one can conclude
\begin{equation}
    \|T_\rho f\|_q \leq \|f\|_p \leq \|f\|_1^{1-\theta} \|f\|_q^{\theta}, \quad \text{where} \quad \frac{1}{p} = 1-\theta + \frac{\theta}{q} \ \iff\  \theta = \frac{1-\tfrac{1}{p}}{1-\tfrac{1}{q}}.
\end{equation}
This shows Samorodnitsky's inequality holds with $\theta = 1 - \frac{1-\rho^2}{1+(q-1)\rho^2}$.  However this is
always strictly larger than the optimal value we know from \Cref{thm:main}, namely $\lambda(q,\tfrac12,\rho) = \frac{1}{q-1} \log_2\left(\frac12(1+\rho)^q + \frac12(1-\rho)^q\right)$.
This may justify the fact that  Samorodnitsky's determination~\cite{Sam20} of the optimal value of $\lambda(q,\tfrac12,\rho)$, even just for integer~$q \geq 2$, appears more difficult than the proof of the ``two-point inequality'' needed for the hypercontractive inequality.

\paragraph{General probability spaces.}
We continue the comparison in the general case, beyond the uniform distribution on $\mathbb{F}_2$.
Here, we remark that the optimal hypercontractive inequality for general finite probability spaces is known in the particular case of $p = 2$. 
The proof comprises a very technical paper of Lata{\l}a and Oleszkiewicz~\cite{LO04} for the $|\Omega| = 2$ case, together with a general reduction to that case by Wolff~\cite{Wol07}.  When $p \neq 2$, the optimal hypercontractive inequality for general finite probability spaces is not known. 
We bring up these facts as an
attempted explanation  
for the fact that our proof  of \Cref{eqn:generaln1} (in \Cref{sec:general}), in the case of general probability spaces $(\Omega,\mu)$ and general real $q \in [2,\infty]$, $\rho \in (0,1)$, is somewhat complicated. 
We do, however, give simplified proofs in the boundary  cases of $q = 2, \infty$; these are also the most interesting cases for known applications.

\paragraph{The induction.}
To conclude our comparison with hypercontractivity, we observe a rather striking aspect of Samorodnitsky's inequality. Namely, the ``general-$n$'' form in~\eqref{eqn:generaln} does not immediately resemble the $n = 1$ case from~\eqref{eqn:generaln1}, as in the hypercontractivity inequalities. Indeed, Samorodnitksy's insight to derive the general-$n$ case seems quite non-obvious, although in the end his inductive proof is modeled after that of hypercontrativity.  In~\Cref{sec:induction}, we give an equivalent form of Samorodnitsky's inequality: 
\begin{equation}
    \|T_\rho f\|_{x:q} \leq \|T_\sigma f\|_{x:q,\sigma:0}\;.
\end{equation}
Above, $\sigma$ on the right hand side is a random
$\Ber(1/2)$ i.i.d.~vector, and the expression
$\|\cdot\|_{x:q,\sigma:0}$
means that first (inside) the $q$-norm is taken with respect
to $x$ and then (outside) the $0$-norm, i.e., geometric mean,
with respect to $\sigma$.
We then give a general 
proof that unifies the induction step for several known 
inequalities. In particular, it gives as an immediate corollary
the induction steps for generalized Samorodnitsky's inequality and for hypercontractivity,
as well as for reverse hypercontractivity~\cite{MOS13}
and the ``randomization-symmetrization'' inequalities used in the proof
of the sharp threshold theorem by Bourgain~\cite{FB99}.
After introducing some notation, the only tool used in the proof
is a general form of the Minkowski inequality.

\paragraph{Strong data processing inequalities.}
Recall that for two probability distributions $\mu,\nu$
on a set $\Omega$, the R{\'e}nyi divergence
of order $q$ is given by
$D_q(\nu||\mu)\coloneq\frac{1}{q-1}
\log\sum_{x\in\Omega}\frac{\nu(x)^q}{\mu(x)^{q-1}}$.
The $n = 1$ case of our main \Cref{thm:main}
has an equivalent natural formulation in terms of R{\'e}nyi divergences. The following theorem follows from \Cref{thm:general-q-norm} and \Cref{rem:sdpi}.
\begin{theorem}
\label{thm:base-case-intro}
Let $\mu$ be a (full-support) probability distribution on finite space $\Omega$.  Then for all $q \in [2,\infty]$ and $\rho \in (0,1)$, the following holds:  If $\nu$ is another distribution on $\Omega$, and $\nu T_\rho$ denotes the mixture $\rho \nu + (1-\rho)\mu$, then 
\begin{equation}
D_q(\nu T_\rho||\mu T_\rho)
=D_q(\nu T_\rho||\mu)
\le\lambda  D_q(\nu||\mu)\;,
\end{equation}
where $\lambda = \lambda(q,\mu^*,\rho)$ is the (optimal) constant from \Cref{eqn:optimallam}.
The inequality is  tight if and only if
$\nu=\mu$ or $\nu(x^*)=1$ for some $x^*$
with $\mu(x^*)=\mu^*$.
\end{theorem}

In the language of information theory, 
$\lambda$ is the \emph{strong data processing inequality (SDPI) constant} for the R{\'e}nyi
divergence of order $q$, distribution $\mu$, and
probability kernel $T_\rho$; see \Cref{rem:sdpi} for further discussion.

\subsection{Applications to error-correcting codes}
In Samorodnitsky's work~\cite{Sam19,Sam20}, he used his inequality to establish several new results in the theory of binary error-correcting codes. These applications were taken further in~\cite{HSS21}, which showed
that the $q=2$ case of the inequality implies bounds on the weight distribution of a binary linear error-correcting code in terms of its conditional entropy on the binary erasure channel (BEC).
As a corollary, \cite{HSS21} showed that (under a very mild distance assumption), any binary linear code with vanishing
decoding error probability on BEC$(\lambda)$ also has vanishing decoding error probability on all binary memoryless symmetric channels with capacity above some
$C=C(\lambda)$ (under optimal MAP decoding, albeit not necessarily with an efficient decoder).
This gives an automatic method for converting positive results about code performance on the BEC
into results on other channels, including the binary
symmetric channel (BSC); in particular, no assumptions
on the symmetry group of a code are required.
As discussed in \cite{HSS21}, one may think of this as showing that there is a universal notion of a binary code being ``good'' in Shannon's model (independent of the choice of channel), complementing the notion of ``good'' codes in Hamming's model.

\Cref{thm:main} lets us extend these consequences to linear codes over any finite alphabet.  First, we derive a bound on the weight distribution of any linear code:
\begin{theorem}\label{thm:weight-distribution-intro}
    Let $k$ be a prime power,  $0 \leq \lambda \leq 1$, and write $z = z_k(\lambda) = \frac{k^\lambda - 1}{k-1}$ as in \Cref{eqn:zk}.
    Then if $\cC\subseteq \mathbb{F}_k^n$ is any linear code, its weight distribution $(a_0,\ldots,a_n)$ (where $a_i = \#\{x \in \cC : |x| = i\}$) satisfies
    \begin{align}
        \log_2 \sum_{i=0}^n a_i \cdot z^i  \leq H(X|Y)\;,
    \end{align}
    where $H(X|Y)$ denotes the base-2 conditional entropy 
    of the input $X$ drawn uniformly from $\cC$ given noisy output~$Y$,
    which is the result of transmitting $X$ through
    $\kEC$ 
    (the $k$-ary erasure channel with erasure probability
    $\lambda$).
\end{theorem}

Then, extending \cite{Sam19,HSS21}, we obtain the following corollary:
\begin{corollary}
\label{thm:kec-to-ksc}
    Let $k$ be any prime power and let $\{\cC_n\}_{n}$ be family of linear codes, $\cC_n \subseteq \mathbb{F}_k^n$, with minimum distance
    lower bound $d(\mathcal{C}_n)\ge \omega(\log n)$.
    Suppose, for some $0 < \lambda < 1$, that $\cC_n$ has vanishing decoding error probability\footnote{
    In fact, if this error probability is $o(1/n)$, the minimum
    distance assumption can be dropped.
    } 
    on $\kEC(\lambda)$.
    Then $\cC_n$ also has vanishing decoding error probability on \emph{any} channel $W:\mathbb{F}_k\to\mathcal{Y}$ satisfying
    $\Bh(W) < z_k(\lambda)$, where $\Bh(W)$ denotes the ``Bhattacharyya coefficient'' of~$W$.
\end{corollary}
Here $\Bh(W)$ is a measure of the ``noisiness'' of channel~$W$, which is well studied, at least in the case of binary memoryless symmetric
(BMS) channels
(its formula is given in \Cref{def:bhatt}). 
In particular $\Bh(W) < 1$, except in the
case where the channel has two completely confusable input symbols (i.e., there are $x \neq x' \in \mathbb{F}_k$ with $W({\cdot} | x)$ and $W({\cdot} | x')$ identical).

As an example corollary, we can apply the above to the $k$-ary symmetric channel $\kSC(\eta)$ with error parameter~$\eta$,
i.e., the channel which keeps the original symbol
with probability $\eta$ or outputs any other symbol
with probability $\eta/(k-1)$. Its (base-$k$) Shannon capacity
$C(\eta)$ and its
Bhattacharyya coefficient are easy to compute. Working out the asymptotics leads to the following conclusion:
\begin{corollary}   \label{cor:ksc}
    Fix a prime power~$k$.
    Then there is a certain function $\delta = \delta(\eta) \sim  C(\eta)/2$ (as $\eta \to 0$) such that the following holds:
    If a family of linear codes $\cC_n \subseteq \mathbb{F}_k^n$ has $d(\cC_n) \geq \omega(\log n)$ and vanishing decoding error probability on $\kEC(1-\delta)$, then it also has vanishing decoding error probability on $\kSC(\eta)$.
\end{corollary}

\paragraph{Undetected error probability.}
Another work by Samorodnitsky~\cite{Sam22} shows that the
$q=\infty$ case of his inequality implies an upper bound on the
\emph{undetected error probability} of a binary linear code in terms
of conditional entropy on the BEC.
In this work, in \Cref{sec:pue}, we extend Samorodnitsky's result to linear codes over any finite field, with an alternative proof that uses the $q=2$ case of \Cref{thm:main}, rather than $q=\infty$.

\paragraph{Other applications in coding theory.}  We close by mentioning other applications of Samorodnitsky's inequalities
and other related inequalities.
In \cite{Sam22}, Samorodnitsky uses the  $q=4$ case of the inequality to study $4$-tuples of codewords which satisfy a certain
distance property. This is relevant to his conjecture
in~\cite{Sam23}.
Other inequalities (for example, inequalities for Shannon
entropy which can be interpreted as a limiting case $q\to 1$)
have been
used to establish consequences for the most informative boolean function conjecture~\cite{Sam16}, the
entropy rate of hidden binary Markov processes~\cite{Ord16, KH24}, list decoding 
on the BSC~\cite{Haz23, KH24},
analysis of the linear programming 
bound for binary codes~\cite{LS22}, and
achieving secrecy on the binary wiretap 
channel~\cite{PB23}.

\subsection{Proof strategy for \Cref{thm:main}}

We highlight an interesting technical aspect
of our proof of \Cref{thm:main}.
For several proof strategies that we tried,
even though the optimal value of $\lambda$
is determined by an ``extremal'' dictator function,
we encountered other local extrema which seemed
difficult to handle.
Ultimately, we proceed by proving a more general
statement concerning bounded random variables.
For a random variable $X$, let
$R_{q,\rho}(X)\coloneqq\frac{\ln \EE(1+\rho X)^q}{\ln \EE(1+X)^q}$. Our main technical theorem (which is restated and proven as
\Cref{thm:general-q-rv} in \Cref{sec:proof-general-q}) reads:

\begin{theorem}
\label{thm:general-q-intro}
Let $q\ge 2$, $0<\rho<1$, and $\alpha>0$.
Let $X$ be any finitely supported random
variable such that $\EE X=0$, $-1\le X\le \alpha$,
and $X\not\equiv 0$.
Let $X_\alpha$ be
a random variable that takes value $\alpha$ with probability $1/(1+\alpha)$ and $-1$ with probability
$\alpha/(1+\alpha)$. Then,
\begin{align}
\label{eq:67-intro}
R_{q,\rho}(X)\le R_{q,\rho}(X_\alpha)\;,
\end{align}
with equality if and only if $X$ and $X_\alpha$
are equal in distribution.
\end{theorem}

It is easy to show that \Cref{thm:general-q-intro}
implies \Cref{thm:main} and we do so
in \Cref{sec:proof-general-q}.
As opposed to \Cref{thm:main},
the more general \Cref{thm:general-q-intro}
can be proved in a quite
natural (though still complicated) way.
First, we show that the supremum of
$R_{q,\rho}(X)$ can only be attained by a random
variable with binary support.
Then, we show that
for a fixed $\alpha>0$, if $\supp(X)=\{y,\alpha\}$,
then $R_{q,\rho}(X)$ is maximized if
$\supp(X)=\{-1,\alpha\}$. Finally,
we show that $R_{q,\rho}(X_\alpha)$ is an increasing
function of $\alpha$.

\subsection{Open directions}
We leave several open questions for future work:
\begin{itemize}
    \item We do not study the case of $1<q< 2$; 
    in fact in this range of $q$ we are not aware of
    \emph{any} result for non-binary alphabets.
    In general, (as already implied by~\cite{Sam19}),
    in this case one should not expect that
    the optimal $\lambda$ is determined by a dictator
    function.
    \item For non-binary spaces, it also makes sense to consider some values of $\rho<0$.  This range of $\rho$ seems to behave differently, as we demonstrate for $q=\infty$ in~\Cref{thm:tensorization-inf-norm}. 
    \item Recall \Cref{thm:base-case-intro}, which states the base case of \Cref{thm:main} in terms of an SDPI for R{\'e}nyi entropies for the specific operator $T_\rho$. We leave open if some of our ideas can be applied more broadly for other operators.
    \item Finally, the general scheme we give in \Cref{sec:induction} might be used as an inspiration in the search for new inequalities.
\end{itemize}

\subsection{Outline of the paper}
\begin{itemize}
    \item \Cref{sec:induction} gives a unified induction/tensorization proof that can be used to generalize the $n = 1$ case of  Samorodnitsky's inequality, hypercontractivity, reverse hypercontractivity, and other inequalities.
    \item \Cref{sec:general} is devoted to proving the $n = 1$ base case of our \Cref{thm:main}.
    In particular, \Cref{thm:main} follows from \Cref{thm:tensorization-q-norm}
    and~\Cref{thm:general-q-norm} there.
    \item \Cref{sec:coding} establishes our main coding-theoretic applications, including \Cref{thm:weight-distribution-intro}
    (see \Cref{decoding_thm}), \Cref{thm:kec-to-ksc}
    (see \Cref{cor:kec-to-ksc}), and 
    \Cref{cor:ksc} (seee \Cref{cor:asymm}).
    \item \Cref{sec:pue} contains our result on the probability of undetected errors on $k$-ary symmetric channels.  
    \item The appendices contain some miscellaneous calculations.
\end{itemize}

\section{General (hyper)contractivity induction}
\label{sec:induction}

In this section we give a unified tensorization proof that can be applied
in several settings including boolean hypercontractivity, reverse hypercontractivity,
and the inequalities of Samorodnitsky. We need to set the stage
and introduce some notation first.

Let $(X_i)_{i=1}^n$ be independent random variables such
that $X_i$ is sampled from a finite, fully supported
probability space
$(\Omega_i,\mu_i)$.
Let us denote by $(\Omega,\mu)$ the resulting product space.
Let $f\in L^2(\Omega,\mu)$.
Then, let us define inductively the following notation for norms of $f$:
\begin{align*}
\|f\|_{\emptyset} &\coloneqq |f(X_1,\ldots,X_n)|\;,\\
\|f\|_{x_i:q} &\coloneqq \left(\EE_{X_i}|f(X_1,\ldots,X_n)|^q\right)^{1/q}\;.
\end{align*}
Note that both quantities are random variables which are functions of $X_1,\dots, X_n$, %
but $\|f\|_{x_i:q}$ does not depend\footnote{
More precisely, by the notation $\EE X_i$ we mean the conditional expectation
\begin{align*}
\EE_{X_i} f(X_1,\ldots,X_n)\coloneqq\EE\left[f(X_1,\ldots,X_n)\;\vert\;
X_1,\ldots,X_{i-1},X_{i+1},\ldots,X_n\right]\;.
\end{align*}
}
on $X_i$.
Now, for some sequence $S=(x_{i_1}{:}q_1,\ldots,x_{i_{k}}{:}q_{k})$ 
(where $i_1,\ldots,i_k$ are pairwise distinct) and $j\notin\{i_1,\ldots,i_{k}\}$ let
\begin{align*}
\|f\|_{S,x_{j}:q}
&\coloneqq
\left(\EE_{X_{j}} \|f\|_{S}^{q} \right)^{1/q}\;.
\end{align*}
Furthermore, we extend in the usual way to $q\in\{0,\infty\}$:
\begin{align*}
\|f\|_{S,x_{j}:0}
&\coloneqq
\exp\mleft(\EE_{X_j} \ln \|f\|_{S}\mright)\;,\\
\|f\|_{S,x_{j}:\infty}
&\coloneqq
\max_{X_j} \|f\|_{S}\;.
\end{align*}
In principle, these expressions are valid
for every $-\infty<q\le\infty$.
However,
they are not well defined for $q\le 0$
if $\|f\|_S=0$ with positive probability; in these cases
we adopt the convention 
$\|f\|_{S,x_j:q} \coloneqq0$.

We will abbreviate a contiguous sequence $x_i{:}p,x_{i+1}{:}p,\ldots,x_j{:}p$ to
$x_i^j{:}p$, and also $x_1^n{:}p$ to $x{:}p$.

\begin{definition}
[Noise operator]
Let 
$f\in L^2(\Omega,\mu)$.
For $1 \leq i \leq n$ and $\rho\in\mathbb{R}$, we define an operator 
$T_{i,\rho}$ as
\begin{align*}
T_{i,\rho}f&\coloneqq \rho f+(1-\rho)\EE_{X_i}f\;.
\end{align*}
Later on we might omit the index $i$ if it is clear from the context.
Given a sequence $\rho_1^n = (\rho_1, \dots, \rho_n)$,
we also let
\begin{align}
T_{\rho_1^n}f \coloneqq T_{1,\rho_1}\cdots T_{n,\rho_n} f\;.
\end{align}
and also write   $\rho \coloneqq \rho_1^n$ for short  when
it is clear from the context.
Finally, in line with existing conventions, if $\rho\in\mathbb{R}$
is a constant, then $T_\rho\coloneqq T_{(\rho,\ldots,\rho)}$.
\end{definition}
As $\EE_{X_i}\EE_{X_j} f=\EE_{X_j}\EE_{X_i} f$,
it is easy to check that $T_{i,\rho_i}T_{j,\rho_j} f=T_{j,\rho_j}T_{i,\rho_i} f$.
Consequently, we get the same random variable
regardless of the order of applying $T_{i,\rho}$.
Let us go one step further and allow $\rho_i$ to be a finitely supported random variable.
That is, consider a random vector $\rho_1^n$, with coordinates
independent of $X_1,\ldots,X_n$ and of each other,
such that $\rho_i$ is a finitely supported random variable for each $i$.

Accordingly, we will think of $T_{\rho_1^n} f$ as a function
of $2n$ random variables and consider norms of the form
$\|T_{\rho_1^n} f\|_{x_1^n:p,\rho_1^n:r}$,
which we might shorten for convenience to $\|T_{\rho}f\|_{x:p,\rho:r}$.
With these notations in place, we can give some examples of their use.

\begin{example}
Using this notation, the usual $q$-norm is easily seen to be
\begin{align}
\|f\|_q&=\|f\|_{x_1:q,\ldots,x_n:q}=\|f\|_{x:q}\;.
\end{align}
\end{example}

\begin{example}
\label{ex:samorodnitsky}
The paper by Samorodnitsky~\cite{Sam19} features inequalities for nonnegative functions on the hypercube 
$L^2(\mathbb{F}_2^n,\Ber(1/2)^n)$
that take the form
\begin{align}
\log\|T_\rho f\|_q\le \EE_{S\sim \lambda}\log \|\EE(f|S)\|_q\;.
\label{eq:07}
\end{align}
In that inequality, $S\subseteq [n]\coloneq\{1,\ldots,n\}$ is a random subset where each element
is included independently with probability $\lambda=\lambda(q,\rho)$.
For $S\subseteq[n]$ and $S^c\coloneqq [n]\setminus S$, the expression $\EE(f|S)$ denotes the conditional
expectation operator $\EE(f|S)(x)\coloneqq\EE_{X_{S^c}} f(x_S, X_{S^c})$.

Given $S$, define the vector $\sigma(S)$ as $\sigma(S)_i\coloneqq1(i\in S)$,
and observe that $\EE(f|S)=T_{\sigma(S)}f$. Let $\sigma$ be a random
vector with i.i.d.~$\Ber(\lambda)$ coordinates. Exponentiating both sides
of~\eqref{eq:07}, we can write an equivalent inequality in our notation:
\begin{align}
\|T_\rho f\|_{x:q}\le \|T_{\sigma} f\|_{x:q,\sigma:0}\;.
\label{eq:08}
\end{align}
\end{example}

\begin{example}
\cite[Section 10.4]{OD14} states a ``symmetrized'' inequality
that is then used in the proof of a sharp threshold theorem by Bourgain~\cite{Bou79,FB99}.
In our notation this inequality reads
\begin{align}
\|T_{1/2}f\|_{x:q}\le \|T_{r}f\|_{x:q,r:q}\le\|T_{c_q}f\|_{x:q}\;.
\label{eq:09}
\end{align}
where $0<c_q<1$ is a constant that depends on $q$ and the random vector $r_1^n$ has
i.i.d.~coordinates uniform in $\{-1,1\}$.
\end{example}

We now state the main theorem of this section, which unifies the examples discussed above.
In each case, \Cref{thm:very-general}
allows to obtain
an inequality for functions of $n$~coordinates
from the single-coordinate version.

\begin{theorem}
\label{thm:very-general}
Let $\rho=\rho_1^n$ and $\sigma=\sigma_1^n$ be two random vectors with independent 
coordinates and finite supports.
Furthermore, let $r\le p\le q\le s\le \infty$.

\textbf{Base case.} Assume that for every $1\le i\le n$
and every $f\in L^2(\Omega_i,\mu_i)$ it holds
\begin{align*}
\|T_{\rho_i}f\|_{x_i:q,\rho_i:s} \le \|T_{\sigma_i} f\|_{x_i:p,\sigma_i:r}\;.
\end{align*}

\textbf{Conclusion.}
Then, for every $f\in L^2(\Omega,\mu)$ it holds
\begin{align*}
\|T_{\rho}f\|_{x:q,\rho:s}
&\le
\|T_{\sigma} f\|_{x:p,\sigma:r}\;.
\end{align*}

Furthermore, the theorem also holds if 
both base case and
conclusion are restricted to nonnegative 
functions,
as long as
for every~$i$, every nonnnegative $f\in L^2(\Omega_i,\mu_i)$,
and every possible value of $\rho_i$ and $\sigma_i$,
$T_{\rho_i} f,T_{\sigma_i}f$ are also nonnegative functions.\footnote{
Most typically, this condition
holds if $0\le\rho_i,\sigma_i\le 1$ for every $i$.
We state it somewhat more generally to allow
$\rho_i<0$ in certain cases, and make use of it
in \Cref{thm:tensorization-inf-norm}.
}
\end{theorem}

Before turning to the proof of \Cref{thm:very-general},
let us show how some examples fit into its framework.
The hypercontractive inequality for finite product spaces
(which includes the boolean hypercontractivity
for uniform random bits) reads
$\|T_\rho f\|_q\le \|f\|_p$
for every $f\in L^2(\Omega^n,\mu^n)$, $q>p>1$ and a certain
$\rho=\rho(p,q,\mu)$. 
This can be obtained by applying
\Cref{thm:very-general} with deterministic $\rho_i=\rho$
and $\sigma_i=1$, as well as $r=p$ and $q=s$.
The reverse hypercontractive inequalities for nonnegative functions
first stated in~\cite{Bor82}
and studied in~\cite{MORSS06,MOS13} are of the form
$\|T_\sigma f\|_p\ge \|f\|_q$ for $p<q<1$
and are covered by \Cref{thm:very-general} in a similar manner.
It is also straightforward to confirm that
the inequalities~\eqref{eq:08} and~\eqref{eq:09} stated in the examples
are covered by the theorem.

\begin{example}
For any nonzero $f\in L^2(\Omega_i,\mu_i)$ the function
$t\to\|T_{\exp(-t)}f\|_2^2$ is log-convex in $t\in\mathbb{R}$. This is because
$\|T_{\exp(-t)}f\|_2^2=\exp(-2t)\|f-\EE f\|_2^2+(\EE f)^2$, which is a sum of two
log-convex functions.

By elementary manipulations, it follows that for any positive random variable $\rho$
with geometric mean $\widetilde{\rho}\coloneqq\exp(\EE \ln\rho)$
it holds that $\|T_{\widetilde{\rho}}f\|_2\le \exp\left(\EE_\rho\ln\|T_\rho f\|_2\right)$; or, in our notation,
$\|T_{\widetilde{\rho}}f\|_{x:2}\le \|T_\rho f\|_{x:2,\rho:0}$. 

Let $\rho\coloneqq \rho_1^n$ be a random vector with independent finitely supported
positive coordinates with geometric means
$\widetilde{\rho}_i\coloneqq\exp(\EE\ln\rho_i)$.
Applying
\Cref{thm:very-general}, we obtain
\begin{align}
    \|T_{\widetilde{\rho}} f\|_{x:2}\le \|T_\rho f\|_{x:2,\rho:0}\;.
\end{align}
In other words, the $2$-norm of $T_{\widetilde{\rho}}f$ is at most
the geometric mean of the 2-norms of $T_\rho f$.
This inequality is used in the
paper~\cite{OWZ14}.
\end{example}


We are ready to prove \Cref{thm:very-general}.
The main
technical tool we need is a general form of the Minkowski inequality,
which in our notation reads:

\begin{theorem}[Generalized Minkowski]
\label{thm:minkowski}
Let $f\in L^2(\Omega,\mu)$ and $p\le q\le\infty$,
while $q_1,\ldots,\allowbreak q_{i-1},\allowbreak q_{i+2},\ldots, q_n
\in\mathbb{R}\cup\{\infty\}$.
Then
\begin{align}
\|f\|_{x_1:q_1,\ldots,x_i:p,x_{i+1}:q,\ldots,x_n:q_n}
&\le
\|f\|_{x_1:q_1,\ldots,x_{i+1}:q,x_i:p,\ldots,x_n:q_n}\;.
\end{align}
\end{theorem}
That is, ``pushing outside'' in the ordering variables with smaller $p$
can only increase the norm.
While this is a standard result, for completeness we sketch the proof of this version in \Cref{app:minkowski-proof}.

\begin{proof}[Proof of~\Cref{thm:very-general}]
The proof proceeds by induction on $n$.
Let us start with applying the induction to the first coordinate:
\begin{align}
\|T_{\rho_1^n} f\|_{x_1^n:q,\rho_1^n:s}
&\le\|T_{\rho_1^n} f\|_{x_1:q,\rho_1:s,x_2^n:q,\rho_2^n:s}\label{eq:10}\\
&=\Big\| \|T_{\rho_1} (T_{\rho_2^n}f)\|_{x_1:q,\rho_1:s} \Big\|_{x_2^{n}:q,\rho_2^n:s}
\label{eq:11}
\\&\le \Big\|\|T_{\sigma_1}(T_{\rho_{2}^n)} f\|_{x_1:p,\sigma_1:r}\Big\|_{x_2^n:q,\rho_2^n:s}
\label{eq:02a}\\
&\le\|T_{\rho_2^n} (T_{\sigma_1}f)\|_{x_2^n:q,\rho_2^n:s,x_1:p,\sigma_1:r}\;.
\label{eq:03a}
\end{align}
In the subsequent steps, we apply:
the Minkowski inequality in~\eqref{eq:10},
the definitions of $T_\rho$ and the norm notation in~\eqref{eq:11},
the base case in~\eqref{eq:02a} 
and again the Minkowski inequality in~\eqref{eq:03a}.
Now applying induction on a function of $n-1$ variables (conditioned on $x_1,\sigma_1$):
\begin{align}
\|T_{\rho_2^n} (T_{\sigma_1}f)\|_{x_2^n:q,\rho_2^n:s,x_1:p,\sigma_1:r}
&=
\Big\|\|T_{\rho_2^n} (T_{\sigma_1}f)\|_{x_2^n:q,\rho_2^n:s}
\Big\|_{x_1:p,\sigma_1:r}
\\
&\le
\left\|
{T_{\sigma_1^n}f}
\right\|_{x_2^n:p,\sigma_2^n:r,x_1:p,\sigma_1:r}\label{eq:13}\\
&\le \|T_{\sigma_1^n} f\|_{x_1^n:p,\sigma_1^n:r}\;.
\end{align}
As for the ``furthermore'' statement, it follows since due to
the assumption, if $f$ is nonnegative, then all intermediate
functions in the expressions above are also nonnegative,
and the appplications of induction in~\eqref{eq:02a} and~\eqref{eq:13}
remain valid.
\end{proof}

\section{Noisy function inequalities for general product spaces}
\label{sec:general}
The inequalities of Samorodnitsky~\cite{Sam19,Sam20} are proved for
boolean functions under the uniform input distribution. We prove analogous
inequalities for larger finite alphabets, for arbitrary product
distributions, and for $q \ge 2$. 

\begin{theorem}
\label{thm:tensorization-q-norm}
Let $2\le q<\infty$, $\Omega$ a finite set and $\mu$ a fully supported probability measure
on $\Omega$ with $\mu^*\coloneqq \min_{x\in\Omega}\mu(x)$. Let $0\le\rho\le 1$
and 
\begin{align}
     \lambda=\lambda(q,\mu,\rho)\coloneqq \frac{\ln\left(\mu^*(1 + \rho(\frac{1}{\mu^*}-1))^q + (1 - \mu^*)(1 - \rho)^q \right)}{(q-1)\ln\frac{1}{\mu^*}}
    \;.
\end{align}
Let $\sigma=\sigma_1^n$ be an i.i.d.~random vector with $\Ber(\lambda)$ coordinates.
Then, for every nonnegative $f\in L^2(\Omega^n,\mu^n)$ it holds
\begin{align}
\label{eq:79}
\|T_\rho f\|_{x:q}\le \|T_{\sigma}f\|_{x:q,\sigma:0}\;.
\end{align}
Furthermore, the inequality is tight for $n=1$ and $f$ with exactly
one nonzero value $f(x^*)$ for some $x^*$ 
such that
$\mu(x^*)=\mu^*$.
\end{theorem}

\begin{remark}
If $\mu$ is the uniform distribution on a set of size
$k$, we have
\begin{align}
1-\lambda &= \frac{(q-1)\ln k-\ln \left(\frac{1}{k}(1 + \rho(k-1))^q + \frac{k-1}{k}(1-\rho)^q\right)}{(q-1)\ln k}\\
&=-\frac{1}{q-1}\log_k\left(\left(\rho+\frac{1-\rho}{k}\right)^q+(k-1)\left(\frac{1-\rho}{k}\right)^q\right)\;,
\end{align}
which is equal to 
the Rényi entropy of order $q$
with logarithm base $k$ for a random variable
with probabilities
$\left\{\rho+\frac{1-\rho}k,\frac{1-\rho}{k},\ldots,\frac{1-\rho}{k}\right\}$.
\end{remark}

\begin{remark}
A natural extension of \Cref{thm:tensorization-q-norm}
holds also for differing values of $(\Omega_i,\mu_i,\rho_i)$ with respective
$\lambda_i(q,\mu_i,\rho_i)$. We give a less
 general statement for the sake of readability.
\end{remark}

For the case of $q=\infty$, we prove
the inequality for a broader range of
$-\frac{\mu^*}{1-\mu^*}\le \rho \le 1$.
For $0\le\rho\le 1$, the result follows 
in the limit of finite $q$, but for negative $\rho$ 
we obtain a seemingly different behavior.

\begin{theorem}
\label{thm:tensorization-inf-norm}
Let $\Omega$ be a finite set and $\mu$ a fully supported probability measure
on $\Omega$ with $\mu^*\coloneqq \min_{x\in\Omega}\mu(x)$. 
Let $-\frac{\mu^*}{1-\mu^*}\le\rho\le 1$ and 
\begin{align}
 \lambda(\mu,\rho)&\coloneqq
\begin{cases}
\frac{\ln(\rho\frac{1}{\mu^*}+1-\rho)}{\ln(\frac{1}{\mu^*})}
&\text{if $\rho\ge 0$,}\\
-\frac{\ln(1-\rho)}{\ln(1-\mu^*)}&\text{if $\rho<0$.}
\end{cases}
\end{align}
Let $\sigma=\sigma_1^n$ be an 
i.i.d.~random vector with
$\Ber(\lambda)$ coordinates. Then,
for every nonnegative $f\in L^2(\Omega^n,\mu^n)$ it holds
\begin{align}
\|T_\rho f\|_{x:\infty}
&\le\|T_\sigma f\|_{x:\infty,\sigma:0}\;.
\end{align}
\end{theorem}

\subsection{Base cases and SDPIs}

Due to \Cref{thm:very-general},
\Cref{thm:tensorization-q-norm}
and
\Cref{thm:tensorization-inf-norm} are implied
by the following base case:

\begin{theorem}
\label{thm:general-q-norm}
Let $q \in [2, \infty]$ and $\Omega,\mu,\mu^*,\rho,\lambda$ be as in \Cref{thm:tensorization-q-norm}
or \Cref{thm:tensorization-inf-norm}.
Then, for every nonnegative $f\in L^2(\Omega,\mu)$ it holds
\begin{align}
\label{eq:74}
\|T_\rho f\|_q \leq \|f\|_1^{1- \lambda}\cdot \|f\|^{\lambda}_q \;.
\end{align}
Furthermore, for $0<\rho<1$, the inequality is tight
if and only if $f$ is constant or if it has exactly
one nonzero value $f(x^*)$ such that $\mu(x^*)=\mu^*$.

For $\rho<0$ and $q=\infty$, the inequality
is tight if and only if $f$ is constant
or if $f(x^*)=0$ for some $x^*$ such
that $\mu(x^*)=\mu^*$ and $f(x)$ is
otherwise constant for $x\ne x^*$.
\end{theorem}

By standard manipulations,
both \Cref{thm:tensorization-q-norm}
and \Cref{thm:tensorization-inf-norm}
follow from \Cref{thm:general-q-norm}
and~\Cref{thm:very-general}.
For $-\frac{\mu^*}{1-\mu^*}\le\rho<0$ and $q=\infty$, one observes that 
\begin{align}
T_\rho f(x)&=
\big(\mu(x)+\rho(1-\mu(x))
\big)f(x)
+\sum_{y \ne x}(1-\rho)\mu(y)f(y)\\
&\ge \left(
\mu(x)-\frac{\mu^*}{1-\mu^*}(1-\mu(x))
\right)f(x)
+\sum_{y \ne x}(1-\rho)\mu(y)f(y)
\ge 0\;,
\end{align}
so $T_\rho f$ is nonnegative whenever $f$ is nonnegative,
therefore \Cref{thm:very-general} is applicable.

As both sides of~\eqref{eq:74} are homogenous with respect
to multiplication by a nonnegative constant, it suffices to prove \Cref{thm:general-q-norm}
restricted to the case $\|f\|_1=1$.

\begin{remark}
\label{rem:sdpi}
Let $\nu,\mu$ be two probability distributions over
$\Omega$ such that $\mu$ is fully supported.
Recall the formula for Rényi divergence
$D_q(\nu||\mu)=\frac{1}{q-1}\log\sum_{x\in\Omega}
\frac{\nu(x)^q}{\mu(x)^{q-1}}$. Given a Markov kernel
$K$, one defines the \emph{strong data processing
inequality (SDPI)
constant}
\begin{align}
\label{eq:70}
\eta_q(\mu,K)\coloneqq
\sup_{\nu:\nu\ne\mu}\frac{D_q(\nu K||\mu K)}{D_q(\nu||\mu)}\;.
\end{align}
There is a broad literature studying
SDPIs for various divergences,
see~\cite[Chapter 33]{PW25} generally and
\cite{Sas19, JEG24, GSOS25} in the context
of Rényi divergences.

There is a natural bijection between distributions
over $\Omega$ and nonnegative functions in
$L^2(\Omega,\mu)$ with $\|f\|_1=1$, given by
$f_\nu(x)\coloneqq\frac{\nu(x)}{\mu(x)}$.
Furthermore, it holds 
$T_\rho f_\nu=f_{\nu T_\rho}$, where\footnote{
This notation is somewhat misleading as $T_\rho$
depends implicitly on $\mu$.
}
$\nu T_\rho\coloneqq\rho\nu+(1-\rho)\mu$.
Since we can check that
\begin{align}
\label{eq:80}
\log \|f_\nu\|^q_q  
&=(q-1)D_q(\nu||\mu)\;,
\end{align}
\eqref{eq:74} is equivalent to
\begin{align}
D_q(\nu T_\rho||\mu T_\rho)=D_q(\nu T_\rho ||\mu)\le\lambda D_q(\nu||\mu)\;,
\end{align}
in particular
\Cref{thm:general-q-norm} implies \Cref{thm:base-case-intro} and the fact that
$\eta_q(\mu,T_\rho)=\lambda(q,\mu,\rho)$.

This seems relevant to some of the results
in~\cite{JEG24}; for example their Theorem~4 says
that if $\eta_2(\mu,K)<1$, then the supremum
in~\eqref{eq:70} is achieved for a distribution with
less than full support. In our case of $K=T_\rho$
and general $q\ge 2$, the supremum is achieved by
an extreme case of the distribution with support size
one.

More generally, we can consider the bijection
$f_\nu(x)=\frac{\nu(x)}{\mu^n(x)}$ between probability
distributions $\nu$ on $\Omega^n$ and $f\in L^2(\Omega^n,\mu^n)$
with $\|f\|_1=1$. Then, it holds
$T_\rho f_\nu=f_{\nu T_\rho}$, where
$\nu T_\rho$ is the distribution obtained by first sampling
from $\nu$, and subsequently, for every coordinate independently,
either keeping the original coordinate with probability $\rho$,
or resampling it from $\mu$ with probability $1-\rho$.
Furthermore, if we consider $\EE(f_\nu|S)$ as an element
of $L^2(\Omega_S,\mu_S)$, then we have
$\EE(f_\nu|S)=f_{\nu_S}$, i.e., conditional expectations
of $f_\nu$ correspond to marginal probability distributions.
Then, applying~\eqref{eq:80}, \eqref{eq:79} is equivalent to
\begin{align}
D_q(\nu T_\rho||\mu^n T_\rho)
=D_q(\nu T_\rho||\mu^n)
\le \EE_{S\sim\lambda} D_q(\nu_S||\mu_S)\;.
\end{align}
See \cite[Section 6]{PY17}, for a similar point of
view in the context of Shannon entropy.
\end{remark}

\paragraph{Plan for proving \Cref{thm:general-q-norm}}

In \Cref{sec:proof-2-norm}, we prove \Cref{thm:general-q-norm}
restricted to $q=2$. This case is easier to prove than
the general one,
it is the most fruitful for currently known
applications, in particular it is the only case 
required for \Cref{sec:coding}
and \Cref{sec:pue}.
Then, in \Cref{sec:proof-inf-norm} we give two
self contained proofs for $q=\infty$.
We include a proof for $\rho\ge 0$ due to its
simplicity, and a separate proof for $\rho<0$ as this
case requires a different argument.
Finally, \Cref{sec:proof-general-q} contains a
proof of the remaining general case $2<q<\infty$.

\subsection{Proof of \Cref{thm:general-q-norm}
for $q=2$}
\label{sec:proof-2-norm}

Let $\psi(x)\coloneqq x\ln x$ continuously extended to
$\psi(x)=0$. Recall that $\psi$
is strictly convex for $x\ge 0$.
Let us start with two lemmas that involve this
function.

\begin{lemma} \label{lem-conv}
Let $0\le \rho \le 1$, $x\ge -1 $. Then 
\[ \rho\psi(1+x)\ge \psi(1+\rho x)\;. \]
Furthermore, the inequality is strict if
$0<\rho<1$ and $x\ne 0$.
\end{lemma}

\begin{proof}
By Jensen's inequality $\psi(1+\rho x)=
\psi(\rho \cdot(1+ x) + (1-\rho)\cdot 1) \le \rho \psi(1+x) + (1-\rho) \psi(1) = \rho \psi(1+x).$ Furthermore, by the strict convexity of $\psi$ the inequality
is strict under the additional conditions.
\end{proof}

\begin{lemma}
\label{lem:lambda-rho-decreasing}
Let $0<\rho<1$. The function $\lambda_\rho(k)\coloneqq \frac{\ln\left(\rho^2 k+1-\rho^2\right)}{\ln k}$ is strictly increasing for $k\ge 1$.
\end{lemma}

\begin{proof}
As the derivative is
\begin{align}
\lambda'_\rho(k)
&=\frac{\rho^2k\ln k-(\rho^2k+1-\rho^2)\ln\left(\rho^2 k+1-\rho^2\right)}{(\rho^2 k+1-\rho^2)k\ln^2 k}
&=\frac{\rho^2\psi(k)-\psi(\rho^2k+1-\rho^2)}{(\rho^2 k+1-\rho^2)k\ln^2 k}
\;,
\end{align}
it is sufficient to show
that the numerator is strictly positive; that is,
$\rho^2\psi(k)>\psi(\rho^2k+1-\rho^2)$ for every $k> 1$.
But that is true by \Cref{lem-conv} applied
for $\rho=\rho^2$ and $x=k-1$.
\end{proof}

\begin{proof}[Proof of \Cref{thm:general-q-norm}, $q=2$]
For $\rho\in\{0,1\}$, the statement is easy to check, so we henceforth assume
$0<\rho<1$. Furthermore, by homogeneity we
can assume without loss of generality that $\|f\|_1 =1$. It is easy to check
that for a nonnegative $f$ it also holds
$\|T_\rho f\|_1=\|f\|_1=1$.

Since $\|f\|_1=1$, by the Cauchy--Schwarz inequality it holds
$\|f\|_2\ge 1$, with equality if and only if $f=1$.
As~\eqref{eq:74} is satisfied for $f=1$ and every $0\le \lambda\le 1$,
establishing~\eqref{eq:74} is reduced to checking
that
\begin{align}
\label{eq:81}
Q(f)\coloneqq \frac{\ln \|T_\rho f\|_2}{\ln \|f\|_2}\le\lambda=\lambda(2,\mu,\rho)\;,
\end{align}
for every nonconstant $f$ with $\|f\|_1=1$.
Furthermore, finding all nonconstant
$f$ which satisfy~\eqref{eq:74} with equality is equivalent
to finding all nonconstant $f$ with  $Q(f)=\lambda$.

Fix some $x_1\ne x_2\in\Omega$ and let us write $\mu_i\coloneqq \mu(x_i)$ for $i=1,2$. 
Let us also fix values of
$f(x)$ for $x\in\Omega\setminus\{x_1,x_2\}$. Letting 
$a\coloneqq \frac12(1-\sum_{x\ne x_1,x_2}\mu(x)f(x))$, the realizable values of
$(f(x_1),f(x_2))$ such that $f$ is nonnegative and $\|f\|_1=1$ are
\begin{align}
\left\{\left(\frac{a+y}{\mu_1},\frac{a-y}{\mu_2}\right): -a \le y\le a
\right\}\;.
\end{align}

We will now consider function $Q(y)$ which is defined
as $Q(f)$ for $f$ with $f(x_1)=\frac{a+y}{\mu_1},f(x_2)=\frac{a-y}{\mu_2}$
and with previously fixed $f(x)$ for $x\notin\{ x_1,x_2\}$. We are going to show that
$Q(y)<\max(Q(-a),Q(a))$ for every valid choice of $-a<y<a$. If $a=0$ this holds vacuously,
so let us assume $a>0$. Let
\begin{align}
v_2&\coloneqq \|f\|^2_2=\mu_1\left(\frac{a+y}{\mu_1}\right)^2
+\mu_2\left(\frac{a-y}{\mu_2}\right)^2
+\sum_{x\ne x_1,x_2}\mu(x)f(x)^2\;,\\
v_1&\coloneqq \|T_\rho f\|_2^2
=\mu_1\left((1-\rho)+\rho\frac{a+y}{\mu_1}\right)^2
+\mu_2\left((1-\rho)+\rho\frac{a-y}{\mu_2}\right)^2\\
&\qquad\qquad\qquad\qquad+\sum_{x\ne x_1,x_2}\mu(x)(1-\rho+\rho f(x))^2\\
&=1-\rho^2+\rho^2 v_2\;.
\end{align}
The derivative of $Q(y)$ is given by
\begin{align}
Q'(y) &= \frac{\partial}{\partial y}\frac{\ln v_1}{\ln v_2}
=\frac{\frac{1}{v_1}\cdot \frac{\partial v_1}{\partial y}\cdot \ln v_2
-\ln v_1\cdot\frac{1}{v_2}\cdot \frac{\partial v_2}{\partial y}}{\ln^2 v_2}
=\frac{\frac{\partial v_2}{\partial y}\cdot\left(\rho^2\psi(v_2)-\psi(v_1)\right)}{v_1\cdot v_2\cdot \ln^2 v_2}\\
&=\frac{2\big((\mu_2-\mu_1)a+(\mu_2+\mu_1)y\big)\cdot\left(\rho^2\psi(v_2)-\psi(v_1)\right)}{\mu_1\mu_2v_1 v_2 \ln^2 v_2}\;.
\end{align}
If there exists $x\ne x_1,x_2$ such that $f(x)\ne 1$, then $v_2>1$ and,
as $\rho^2\psi(v_2)=\rho^2\psi(1+v_2-1)>\psi(1+\rho^2 v_2-\rho^2)=\psi(v_1)$
by \Cref{lem-conv}, the sign of $Q'(y)$ is equal to the sign
of linear function with positive slope $(\mu_2-\mu_1)a+(\mu_2+\mu_1)y$, which has its only zero
at $y_0\coloneqq \frac{\mu_1-\mu_2}{\mu_1+\mu_2}a$.
As it is easy to see $-a< y_0< a$, it follows that
$Q(y)$ is strictly decreasing from $-a$ to $y_0$ and strictly increasing
from $y_0$ to $a$. In particular, $Q(y)<\max(Q(-a),Q(a))$ for every $-a<y<a$.

If $f(x)=1$ for every $x\ne x_1,x_2$, then the analysis
is the same except that for $y=y_0$ it holds
$v_1=v_2=1$ and $Q(y_0)$ is ill defined.
However, it still holds that
$Q(y)$ is decreasing for $-a<y<y_0$ and increasing
for $y_0<y<a$ and consequently $Q(y)<\max(Q(-a),Q(a))$
for every $y\ne y_0$.

To sum up, we proved $Q(y)<\max(Q(-a),Q(a))$ for every choice of
$x_1,x_2$, $f(x)$ for $x\notin\{x_1,x_2\}$, and $-a<y<a$. Applying this property
repeatedly, it is easy to see that maximum of $Q(f)$ can only be attained
for $f$ which has exactly one non-zero value.

Accordingly, for $x\in\Omega$, let us define 
$f_x\in L^2(\Omega,\mu)$ as $f_x(x)\coloneqq 1/\mu(x)$ and $f_x(x')\coloneqq 0$ for
$x'\ne x$. It is easy to check that $Q(f_x)=\lambda_\rho(1/\mu(x))$.
By \Cref{lem:lambda-rho-decreasing}, the maximum of those values
is $Q(f_{x^*})$ for any $x^*$ such that $\mu(x^*)=\mu^*$.
As $\lambda(2,\mu,\rho)=\lambda_\rho(1/\mu^*)$, this establishes both~\eqref{eq:81}
and the tightness condition.
\end{proof}

\subsection{Proof of \Cref{thm:general-q-norm}
for $q=\infty$}
\label{sec:proof-inf-norm}

\begin{proof} [Proof of \Cref{thm:general-q-norm} for $q = \infty$, $0\le \rho\le 1$]
As in the case of $q=2$, let us assume that $0<\rho<1$ and $f$ is a nonconstant function
with $\|f\|_1=1$. Checking~\eqref{eq:74} is now reduced to
showing
\begin{align}
\label{eq:12}
Q(f)&\coloneqq \frac{\ln \|T_\rho f\|_\infty}{\ln \|f\|_\infty}
\le\lambda=\lambda(\infty,\mu,\rho)
\end{align}
for such functions, and $\eqref{eq:74}$ is tight if and only if 
$Q(f)=\lambda$.
Let $y\coloneqq \max_{x\in\Omega} f(x)$.
Due to the assumption that $f$ is not constant note that
$y>1$.
Then, it holds
$\|f\|_\infty=y$ and $\|T_\rho f\|_\infty=\rho y+1-\rho$.
Therefore, \eqref{eq:12} is equivalently written
as
\begin{align}
Q(f)=Q(y)=\frac{\ln(\rho y+1-\rho)}{\ln y}\le \lambda\;.
\end{align}
The derivative $Q'(y)$ is given by
\begin{align}
\label{eq:15}
Q'(y)&=\frac{\rho y \ln y - (\rho y + (1-\rho)) \ln(\rho y+1-\rho)}{y(\rho y + 1-\rho) (\ln y)^2}
\end{align}
Recall that $\psi(x)=x\ln x$. The numerator in~\eqref{eq:15} can be 
rewritten as
\begin{align}
\rho\psi(1+(y-1))-\psi(1+\rho(y-1))>0\;,
\end{align}
where the inequality follows by \Cref{lem-conv}.
Hence, $Q'(y)>0$ for $y>1$ and $Q(y)$ is strictly increasing.
Since $f\ge 0$ and $\|f\|_1=1$, 
the maximum possible $y$ is $y=\frac{1}{\mu^*}$ for $f$
with one non-zero value $f(x^*)$ at $x^*$ such that
$\mu(x^*)=\mu^*$. Hence, we have
\begin{align}
Q(f)&\le Q\left(\frac{1}{\mu^*}\right)
=\lambda\;,
\end{align}
and the tightness condition also follows.
\end{proof}

\begin{proof}[Proof of \Cref{thm:general-q-norm} for $q = \infty$, $\frac{-\mu^*}{1-\mu^*}\le \rho < 0$]
If $\rho=-\frac{\mu^*}{1-\mu^*}$, then $\lambda=1$
and $\|T_\rho f\|_\infty\le\|f\|_\infty$, since
\begin{align}
T_\rho f(x)&=\frac{1}{1-\mu^*}\left(
(\mu(x)-\mu^*)f(x)+\sum_{y\ne x}\mu(y)f(y)
\right)\le \|f\|_\infty\;.
\end{align}
Accordingly, let $\rho>\frac{-\mu^*}{1-\mu^*}$.
Furthermore, suppose that $\|f\|_1 = 1$ and that $f$ is not constant.
Let $\ymin:=\min_{x\in\Omega}f(x)$
and $\ymax:=\max_{x\in\Omega}f(x)$.
Then, $\|f\|_\infty =\ymax$ and $\|T_\rho f\|_\infty = 1- \rho + \rho \ymin$.
Note that $0\le \ymin<1<\ymax$.
We wish to show that
\begin{align}
Q(f)&:=
\frac{\ln(1-\rho+\rho\ymin)}{\ln\ymax}\le
\lambda=\lambda(\infty,\mu,\rho)
\end{align}
and determine which functions achieve $Q(f)=\lambda$.

Let $x_0$ be any value such that $f(x_0)=\ymin$. It holds
\begin{align}
\label{eq:54}
\frac{\sum_{x\ne x_0}\mu(x)f(x)}{\sum_{x\ne x_0}\mu(x)}
=\frac{1-\mu(x_0)\ymin}{1-\mu(x_0)}
\ge\frac{1-\mu^*\ymin}{1-\mu^*}\;,
\end{align}
where we used the fact that the expression
$\frac{1-\mu\ymin}{1-\mu}$ is strictly increasing
on $0\le\mu<1$ for every fixed $0\le\ymin<1$.
As the expression on the left hand side of~\eqref{eq:54}
is a weighted mean of values of $f(x)$, it follows that
$\ymax\ge\frac{1-\mu^*\ymin}{1-\mu^*}$, and that
\begin{align}
Q(f)\le\frac{\ln(1-\rho+\rho\ymin)}{\ln\left(\frac{1-\mu^*\ymin}{1-\mu^*}\right)}\;.
\label{eq:55}
\end{align}
Furthermore, \eqref{eq:55} is tight if and only if
$\ymax=\frac{1-\mu^*\ymin}{1-\mu^*}$.

We will now show that 
$Q(y):=\frac{\ln(1-\rho+\rho y)}{\ln\left(\frac{1-\mu^*y}{1-\mu^*}\right)}$ is strictly decreasing for $0\le y<1$.
The derivative of $Q$ up to multiplying by a positive function is given by:
\begin{align}\label{exp: neg_rho}
W(y)&:= \rho(1-\mu^* y)\big(\ln(1-\mu^* y) -\ln(1-\mu^*)\big) 
+\mu^*(1-\rho + \rho y)\ln(1-\rho + \rho y)
\end{align}
We need to show that the expression in \eqref{exp: neg_rho} 
is negative. Note that $W(1)=0$. 
To conclude that $W(y) < 0$ for $0\le y<1$,
we will show that $W'(y) > 0$.
Indeed, the derivative is
\begin{align}
W'(y)=\mu^*\rho\left[
\big(\ln(1-\mu^*)-\ln(1-\mu^* y)\big)+\ln(1-\rho+\rho y)
\right]\;.
\end{align}
As $\rho<0$, to show that $W'(y)>0$
it suffices to show that
$\ln\frac{(1-\rho+\rho y)(1-\mu^*)}{1-\mu^*y}<0$ or equivalently
$(1-\rho+\rho y)(1-\mu^*)<1-\mu^*y$.
But this is equivalent to $-\rho(1-\mu^*)<\mu^*$, which
holds since $-\rho<\frac{\mu^*}{1-\mu^*}$.

To sum up, we showed that $Q(y)$ has a unique maximum for $y=0$
and that
$Q(f)\le Q(0)=\lambda$ for every nonconstant $f$.
Furthermore, $Q(f)=\lambda$ if and only if $\ymin=0$ 
and $\ymax=\frac{1}{1-\mu^*}$,
which is only possible if $f(x^*)=0$ for some $x^*$
with $\mu(x^*)=\mu^*$ and $f(x)=\frac{1}{1-\mu^*}$
for every $x\ne x^*$.
\end{proof}

\subsection{Proof of \Cref{thm:general-q-norm}
for general $q$}
\label{sec:proof-general-q}

In order to prove \Cref{thm:general-q-norm} for 
general (finite) $q \ge 2$,
it appears more natural to reformulate it 
as a statement about bounded random variables.
For $\alpha>0$ and $k\ge 2$, let
$\mathcal{P}^*_{k,\alpha}$ be the space of all
(distributions of) real random variables $X$
such that $-1\le X\le\alpha$, $\EE X=0$, $X$ is not
identically constant and the
support of $X$ has size at most $k$. Let
$\mathcal{P}^*_\alpha\coloneqq 
\bigcup_{k\ge 2}\mathcal{P}^*_{k,\alpha}$.
For $X\in\mathcal{P}^*_\alpha$,
$q> 1$, and $0\le\rho\le 1$ let
\begin{align}
\label{eq:66}
N_{q,\rho}(X)&\coloneqq \EE(1+\rho X)^q&
R_{q,\rho}(X)&\coloneqq
\frac{\ln N_{q,\rho}(X)}{\ln N_{q,1}(X)}\;.
\end{align}
As $X$ is not constant, it holds
$\EE(1+X)^q>(\EE(1+X))^q=1$, and accordingly
$R_{q,\rho}(X)$
is well defined.

\begin{theorem}
\label{thm:general-q-rv}
Let $q\ge 2$, $0<\rho<1$, and $\alpha>0$. 
Let $X_\alpha$ be
a random variable that takes value $\alpha$ with probability $1/(1+\alpha)$ and $-1$ with probability
$\alpha/(1+\alpha)$. Then, for every 
$X\in\mathcal{P}_\alpha^*$,
\begin{align}
\label{eq:67}
R_{q,\rho}(X)\le R_{q,\rho}(X_\alpha)\;,
\end{align}
with equality if and only if $X$ and $X_\alpha$
have the same distribution.
\end{theorem}

Let us quickly point out that \Cref{thm:general-q-rv}
implies \Cref{thm:general-q-norm}.
\begin{proof}
[\Cref{thm:general-q-rv} implies \Cref{thm:general-q-norm} for $2\le q<\infty$]
Let $f\in L^2(\Omega,\mu)$. The cases of $\rho\in\{0,1\}$
and constant $f$ are easy to handle, so let us assume
$0<\rho<1$, $f$ is not constant and furthermore $\|f\|_1=1$.

Let $Z\sim\mu$ and $X\coloneqq f(Z)-1$. It is easy
to check that $X\in\mathcal{P}^*_\alpha$
for $\alpha=\frac{1}{\mu^*}-1$
and that $R_{q,\rho}(X)=
\frac{\ln\|T_\rho f\|_q}{\ln \|f\|_q}$.
From \Cref{thm:general-q-rv}, it holds
$\frac{\ln\|T_\rho f\|_q}{\ln \|f\|_q}=R_{q,\rho}(X)
\le R_{q,\rho}(X_\alpha)=\lambda(q,\mu,\rho)$.

Furthermore, it can be checked that $X$ has the
distribution of $X_\alpha$ (and hence the inequality is tight) if and only if $f(x^*)>0$ for some
$x^*$ with $\mu(x^*)=\mu^*$ and $f(x)=0$ for $x\ne x^*$.
\end{proof}

\begin{remark} 
Inequality~\eqref{eq:67} holds
for all nonconstant
random variables with $-1\le X\le \alpha$ and $\EE X=0$,
even if the assumption of finite support is dropped.\footnote{
The uniqueness property for 
general random variables seems to require
an additional argument, which we do not include here.
}
For example, given such a random variable $X$, one
can define a sequence $(X_n)$ such that
$X_n\in\mathcal{P}^*_\alpha$ for every $n$ and
$X_n$ converges to $X$ almost surely. Then,
by dominated convergence and by continuity, it holds
$R_{q,\rho}(X)=\lim_{n\to\infty} R_{q,\rho}(X_n)
\le R_{q,\rho}(X_\alpha)$.
\end{remark}

The reformulation of our objective in terms
of \Cref{thm:general-q-rv} seems important.
For several proof strategies we tried
for \Cref{thm:general-q-norm},
we encountered local extrema which we found
difficult to handle. On the other hand, 
the more general \Cref{thm:general-q-rv}
is proved in the following quite natural way,
in three steps.

In~\Cref{sec:reduction-to-binary}, we show
that $\sup_{X\in\mathcal{P}^*_\alpha} R_{q,\rho}(X)$
can only be attained for $X\in\mathcal{P}^*_{2,\alpha}$,
i.e., if $X$ has at most binary support.
In~\Cref{sec:fixed-alpha}, we show that
for a fixed $\alpha>0$, if $\supp(X)=\{y,\alpha\}$,
then $R_{q,\rho}(X)$ is maximized if
$\supp(X)=\{-1,\alpha\}$. Finally,
in~\Cref{sec:fixed-minus-one}, we show that
if $\supp(X)=\{-1,\alpha\}$, then
$R_{q,\rho}(X)$ is an increasing function of
$\alpha$.



\subsubsection{Reduction to the binary case}
\label{sec:reduction-to-binary}



\begin{proposition}\label{thm:binary-reduction}
Let $q\ge 2$, $k\ge 3$, $\alpha > 0$, $0<\rho<1$
and 
$X\in\mathcal{P}^*_{k,\alpha}\setminus 
\mathcal{P}^*_{2,\alpha}$. Then, there exists
$X'\in\mathcal{P}^*_{k,\alpha}$ such that
$R_{q,\rho}(X)<R_{q,\rho}(X')$.
\end{proposition}

\paragraph{Derivative calculation}
Let 
$X\in\mathcal{P}^*_{k,\alpha}\setminus
\mathcal{P}^*_{k-1,\alpha}$ for some $k\ge 3$.
Let $N(\rho)\coloneqq \EE(1+\rho X)^q$
and $Z(\rho)\coloneqq N(\rho)\ln N(\rho)$.

Take some $y,z\in\supp(X)$ such that
$z\notin\{-1,\alpha\}$.
Accordingly, the distribution
of $X$ can be written as
\begin{align}
P_X = p_y\delta_y+p_z\delta_z+\sum_{i=1}^{k-2} p_i \delta_{x_i}\;,
\end{align}
for some $p_y,p_z,p_i>0$. For $\eps$
with $|\eps|< \min(p_y,p_z)$, let $X_{y,z,\eps}$ be the random
variable with distribution
\begin{align}
P_{X_{y,z,\eps}}:=(p_y+\eps)\delta_y
+(p_z-\eps)\delta_{z'}+\sum_{i=1}^{k-2}p_i\delta_{x_i}\;,
\end{align}
where
$z'\coloneqq z+\frac{\eps(z-y)}{p_z-\eps}$.
Then, let $f_{y,z}(\eps)\coloneqq R_{q,\rho}(X_{y,z,\eps})$.

\begin{lemma}\label{lem:fyz-derivative}
In the context above,
let $g(\rho,y,z):=(1+\rho y)^q-(1+\rho z)^q+q\rho\cdot (z-y)(1+\rho z)^{q-1}$.
Then,
\begin{align}
\label{eq:72}
f'_{y,z}(0)=\frac{Z(1) g(\rho,y,z)-Z(\rho)g(1,y,z)}{N(\rho)N(1)\ln^2 N(1)}\;.
\end{align}
In particular,
\begin{align}
\sgn(f'_{y,z}(0))
&=\sgn\big(
Z(1)g(\rho,y,z)-Z(\rho)g(1,y,z)
\big)\;.
\label{eq:46}
\end{align}
\end{lemma}

\begin{proof}
First, note that
$\frac{\mathrm{d}z'}{\mathrm{d}\eps}\rvert_{\eps=0}=\frac{z-y}{p_z}$.
Let $N(\rho,\eps)\coloneqq \EE(1+\rho X_{y,z,\eps})^q$, so that
$N(\rho)=N(\rho,0)$. Then,
\begin{align}
\frac{\mathrm{d}N(\rho,\eps)}{\mathrm{d}\eps}
&=
\frac{\mathrm d}{\mathrm{d}\eps}\left[
(p_y+\eps)(1+\rho y)^q+(p_z-\eps)(1+\rho z')^q
+\sum_{i=1}^{k-2}p_i(1+\rho x_i)^q\right]\\
&=(1+\rho y)^q-(1+\rho z')^q+(p_z-\eps)q\rho(1+\rho z')^{q-1}
\frac{\mathrm{d}}{\mathrm{d}\eps}z'\;,
\end{align}
in particular $\frac{\mathrm{d}N(\rho,\eps)}{\mathrm{d}\eps}
\rvert_{\eps=0}=g(\rho,y,z)$.
Then, \eqref{eq:72} and \eqref{eq:46} follows since
$f_{y,z}(\eps)=\frac{\ln N(\rho,\eps)}{\ln N(1,\eps)}$,
and since by assumptions on $X$
it holds $N(\rho),N(1)>1$.
\end{proof}

\begin{lemma}\label{lem:g-positive}
Let $q\ge 2$, $0<\rho\le 1$ and $y,z\ge -1$. Then, $g(\rho,y,z)\ge 0$ and
furthermore $g(\rho,y,z)=0$ if and only if $y=z$.
\end{lemma}

\begin{proof}
If $y=z$, then it is easy to check that $g(\rho,y,z)=0$. Now, fix $z\ge -1$.
Then, note
\begin{align}
\frac{\partial}{\partial y}g(\rho, y,z)
&=q\rho\big[(1+\rho y)^{q-1}-(1+\rho z)^{q-1}\big]\;.
\label{eq:47}
\end{align}
So, for fixed $0<\rho\le 1$ and $z\ge -1$, $g$ as a function of $y$
is strictly decreasing for $-1\le y<z$ and strictly increasing for $y>z$.
That implies the statement.
\end{proof}

\paragraph{$h$ is increasing}
Let
\begin{align}
h(\rho,y,z)&:=
\begin{cases}
    \frac{g(1,y,z)}{g(\rho,y,z)}&\text{if $y\ne z$,}\\
    \frac{(1+y)^{q-2}}{\rho^2(1+\rho y)^{q-2}}&\text{if $y=z$.}
\end{cases}\;.
\end{align}
A quick calculation using the Taylor expansion shows that
for fixed $z\ge -1$ and $0<\rho<1$,
$h$ is a continuous function of $y\ge -1$.
We state the main technical fact
that will allow a proof of \Cref{thm:binary-reduction}.
\begin{proposition}\label{lem:h-increasing}
For fixed $q > 2, z\ge -1$ and $0<\rho<1$, function
$h(\rho,y,z)$ is strictly increasing
for $y\ge -1$.
\end{proposition}

\begin{proof}[Proof of \Cref{lem:h-increasing}]
Since $h$ is continuous around $y=z$, it is enough to show
that $\frac{\partial}{\partial y} h(\rho,y,z)>0$
for $y\ne z$. As in that case we know
by \Cref{lem:g-positive} that $g(\rho,y,z)\neq 0$, 
this reduces to showing
\begin{align}
\LHS&:=\left(\frac{\partial}{\partial y}g(1,y,z)\right)
\cdot g(\rho,y,z)
-\left(\frac{\partial}{\partial y}g(\rho,y,z)\right)
\cdot g(1,y,z)>0\;.
\label{eq:h-incr}
\end{align}
From~\eqref{eq:47}, we calculate this as
\begin{align}
\LHS
&=
q\left((1+y)^{q-1}-(1+z)^{q-1}\right)
\cdot\left((1+\rho y)^q-(1+\rho z)^q+q\rho(z-y)(1+\rho z)^{q-1}\right)\\
&\qquad -
q\rho \left((1+\rho y)^{q-1}-(1+\rho z)^{q-1}\right)
\cdot\left((1+y)^q-(1+z)^q+q(z-y)(1+z)^{q-1}\right)\;.
\end{align}

If $z =-1$, then $\LHS$ divided by $q(1+y)^{q-1}$
becomes $(1+\rho y)^q - (1-\rho)^q - q\rho(1+y)(1-\rho)^{q-1} - \rho(1+y)(1+\rho y)^{q-1} + \rho(1+y)(1-\rho)^{q-1}$. This expression is zero at $y=-1$ and its derivative with respect
to $y$ is
\begin{align}
(q-1)\rho(1-\rho)\left(
(1+\rho y)^{q-2}-(1-\rho)^{q-2}
\right)>0\;,
\end{align}
therefore indeed $\LHS>0$ for $z=-1$ and $y>z$.

Hence, assume $z>-1$ and $y\ne z$.
Substitute $r\coloneqq q-1,
1+z\coloneqq Z,1+y\coloneqq \alpha Z,
1+\rho z\coloneqq \tilde{Z},
1+\rho y\coloneqq \beta\tilde{Z}$.
Note that if $y>z$ then 
$\alpha=\frac{1+y}{1+z}>
\frac{1+\rho y}{1+\rho z}=\beta>1$,
and similarly if $y<z$, then
$\alpha<\beta<1$.

Using $\rho=\frac{\tilde{Z}(\beta-1)}{Z(\alpha-1)}$, and dividing $\LHS$ by 
$q$ and multiplying by $Z(\alpha-1)^2$,
we need to show
\begin{align}
&Z(\alpha-1)^2((\alpha Z)^r-Z^r)((\beta\tilde{Z})^{r+1}-\tilde{Z}^{r+1}
-(r+1)(\beta-1)\tilde{Z}^{r+1})\\
&\qquad-\tilde{Z}(\alpha-1)(\beta-1)((\beta \tilde{Z})^r-\tilde{Z}^r)
((\alpha Z)^{r+1}-Z^{r+1}
-(r+1)(\alpha-1)Z^{r+1}) > 0\;.
\end{align}
Dividing by $(Z\tilde{Z})^{r+1}$ and rearranging,
we need for $1 < \beta < \alpha$ and for
$\alpha < \beta < 1$,
\begin{align}
\label{eq:73}
(\alpha-1)^2(\alpha^r-1)
(\beta^{r+1}-(r+1)\beta+r)>(\alpha-1)(\beta-1)(\beta^r-1)(\alpha^{r+1}-(r+1)\alpha
+r)\;.
\end{align}
Let $d(\beta)\coloneqq\beta^{r+1}-(r+1)\beta+r$
and $f(\beta)\coloneqq \frac{(\beta-1)(\beta^r-1)}{d(\beta)}$.
As $d(\beta)$ is convex
(recall that $r>1$),
and as $d(1)=d'(1)=0$, we have $d(\beta)>0$ for every
$\beta\ne 1$. Consequently, \eqref{eq:73}
is reduced to showing
$(\alpha-1)f(\alpha)>(\alpha-1)f(\beta)$ for
$1 < \beta < \alpha$ and $\alpha < \beta < 1$.
In turn, this will follow if we show that
$f$ is increasing for $0\le \beta<1$ and for
$\beta>1$, in particular if we show
$f'(\beta) > 0$ for $\beta\ge 0,\beta\ne 1$.
Calculating this and disregarding the positive
denominator $d^2(\beta)$, it holds $f'(\beta)>0$ if
\begin{align}
\beta^{2r}-r^2\beta^{r+1}+2(r^2-1)\beta^r-r^2\beta^{r-1}+1
&=(\beta^r-1)^2-\left(r\beta^{(r-1)/2}(\beta-1)\right)^2\\
&=(\beta^r-r\beta^{(r+1)/2}+r\beta^{(r-1)/2}-1)\\
&\quad \cdot
(\beta^r+r\beta^{(r+1)/2}-r\beta^{(r-1)/2}-1) \\
&=: T_1 \cdot T_2> 0
\end{align}
for $r>1$ and $\beta\ge 0,\beta\ne 1$.
Clearly, $\sgn(T_2)=\sgn(\beta-1)$, so we will be
done if we show $\sgn(T_1)=\sgn(\beta-1)$.
Substitute $\beta\coloneqq\alpha^2$, which gives
$T_1=\alpha^{2r}-r\alpha^{r+1}+r\alpha^{r-1}-1
=\alpha^r(F(\alpha)-F(1/\alpha))$ for 
$F(\alpha)\coloneqq\alpha^r+r/\alpha$.
So we need $F(\alpha)>F(1/\alpha)$ for $\alpha>1$,
or in other words $G(t)>G(-t)$ for
$G(t)\coloneqq F(\exp(t))=\exp(rt)+r\exp(-t)$
and $t>0$.\\
Note that $G'(t)=r(\exp(rt)-\exp(-t))$,
in particular $\sgn(G'(t))=\sgn(t)$.
Furthermore, as the function
$\exp(t)+\exp(-t)$ is increasing for $t>0$,
it follows $G'(t)>-G'(-t)$ for $t>0$, which 
indeed implies $G(t)>G(-t)$ for $t>0$.
\end{proof}

\begin{proof}[Proof of \Cref{thm:binary-reduction}]
We will consider two cases,
$q=2$ and $q>2$. 
In both cases, we will show that there exist
$y,z\in\supp(X)$ such that $z\notin \{-1,\alpha\}$
and $f'_{y,z}(0)\ne 0$, which implies
that for small enough
$\eps$ (positive or negative depending on the sign
of $f'_{y,z}(0)$) it holds
$R_q(\rho,X_{y,z,\eps})>R_q(\rho,X)$.
As it is straightforward to check that $X_{y,z,\eps}\in\cP^*_{k,\alpha}$
for sufficiently small $|\eps|$, this concludes the argument.

For $q=2$, let $y,z\in\supp(X)$ such that
$y\ne z$ and $z\notin\{-1,\alpha\}$.
As $g(\rho,y,z)=\rho^2(y-z)^2$,
by \Cref{lem-conv} it holds
\begin{align}
Z(1)g(\rho,y,z)-Z(\rho)g(1,y,z)
&=(y-z)^2\left(
\rho^2\psi(\EE(1+X)^2)-\psi(\EE(1+\rho X)^2)
\right)
\\&=(y-z)^2\left(\rho^2\psi(1+\EE X^2)-\psi(1+\rho^2\EE X^2)
\right) > 0\;,
\end{align}
therefore by~\eqref{eq:46} $f'_{y,z}(0)>0$.

For $q>2$,
let $z\in\supp(X)$ such that $z\notin\{-1,\alpha\}$
and let $y,y'$ be any two other elements of $\supp(X)$.
By \Cref{lem:h-increasing}, it is not possible that
$h(\rho,y,z)=h(\rho,y',z)=\frac{Z(1)}{Z(\rho)}$.
Accordingly, assume w.l.o.g.~that
$h(\rho,y,z)=\frac{g(1,y,z)}{g(\rho,y,z)}\ne\frac{Z(1)}{Z(\rho)}$.
By~\eqref{eq:46}, that implies
$f'_{y,z}(0)\ne 0$. 
\end{proof}

\subsubsection{The function $R_q(\rho,x_0,y)$ is increasing in $y$}
\label{sec:fixed-alpha}

In light of \Cref{thm:binary-reduction}, 
we can focus on the supremum of $R_{q,\rho}(X)$
over random variables with binary support in $\mathcal{P}_{2,\alpha}$.
To this end, note that since its expectation is zero, such a random variable is
uniquely defined by its two values. Accordingly,
let
\begin{align}
\label{eq:71}
N_q(\rho, x,y)&\coloneqq \frac{y}{y+1}(1+\rho x)^q + \frac{1}{y+1}(1-\rho yx)^q\;,&
R_q(\rho, x, y)&\coloneqq
\frac{\ln N(\rho,x,y)}{\ln N_q(1,x,y)}\;,
\end{align}
and our remaining task is equivalent to finding the
supremum of $R_q(\rho,x,y)$ over the domain
$D_\alpha\coloneqq \{(x,y): 0 < y \leq \frac{1}{x}, 0 < x \leq \alpha\}$. First, we will prove that
for fixed $x_0>0$,
$R_q(\rho,x_0,y)$ is strictly increasing for
$0<y\le 1/x_0$. Then, we will show that
$R_q(\rho,x,1/x)$ is strictly increasing for $x>0$.


\begin{proposition}\label{lem:reduction-to-minus-one}
Let $0<\rho< 1, x_0>0$, $q\ge 2$, and
$R=R(y)\coloneqq R_q(\rho,x_0,y)$.
Then, $R(y)$ is strictly increasing in $0 < y \leq 1/x_{0}$.
\end{proposition}

In the rest of this section we prove \Cref{lem:reduction-to-minus-one}.
Let $N=N(\rho,y)\coloneqq N_q(\rho,x_0,y)$, and
$L\coloneqq \ln N$. Similarly, let 
$\hat{N}\coloneqq N_q(1,x_0,y)$
and $\hat{L}\coloneqq \ln\hat{N}$.
Recall that 
$N>1$ and $L>0$ for $\rho>0$
and $y>0$.
Computing the derivative $R' = \frac{d}{dy} R(\rho,y)$, and omitting the nonnegative denominator gives, for $y>0$,
\begin{equation}
    R' \text{ has the same sign as } L' \hat{L} - L \hat{L}' = \frac{N' \hat{L}}{N} - \frac{L \hat{N}'}{\hat{N}}\;.
\end{equation}
As $R$ is a continuous function of $y$, it is sufficient that
we show
\begin{equation} \label{eqn:switch}
    \frac{N'}{L N} > \frac{\hat{N}'}{\hat{L} \hat{N}}
  \end{equation}
for $0<y< 1/x_0$ and $0<\rho< 1$.

The expression $\frac{N'}{L N}$ is a function of both~$y$ and~$\rho$.
Thinking of $y$ as fixed and $\rho$ as varying, the left and right hand sides of \eqref{eqn:switch} are equal when~$\rho = 1$.
So to establish the strict inequality \eqref{eqn:switch} for $\rho<1$, it suffices to show (for each~$y$) that $\frac{N'}{L N}$ is a strictly decreasing function of~$\rho$.
To this end, we henceforth use subscripted derivative notation on $N = N(\rho, y)$, such as $N_y$ for $\frac{d}{dy} N(\rho, y)$.
Thus we now wish to show,
\begin{equation} \label{eqn:NNN}
    0 > \frac{d}{d\rho} \frac{N_y}{L N} \propto N_{\rho y} L N - (L_\rho N + L N_\rho) N_y
    = L(N_{y \rho} N - N_\rho N_y) - N_\rho N_y.
  \end{equation}
Let us assume for now a claim that will be justified in a short while:
\begin{claim} \label{lem:NNN}
    $N_\rho> 0$ and $N_y> 0$.
\end{claim}
In light of \Cref{lem:NNN} and $L> 0$,
showing \eqref{eqn:NNN} is reduced to showing
\begin{equation}
\label{eq:63}
    L < \frac{N_\rho N_y}{N_{\rho y} N - N_\rho N_y} \quad\iff\quad 0<\frac{M}{W - M}-L,
\end{equation}
where $W \coloneqq N_{\rho y} N$, $M \coloneqq N_\rho N_y$.
With some foresight, we calculate some
partial derivatives, with the substitution $\kappa \coloneqq \rho x_0 \in(0, x_0)$:
\begin{align}
    N_y &= \frac{(1+\kappa)^q - (1-y\kappa)^q}{(1+y)^2} - q\kappa \frac{(1-y \kappa)^{q-1}}{1+y} \\    
    N_\rho &= x_0q  \frac{y}{1+y} \left((1+\kappa)^{q-1} - (1-y\kappa)^{q-1}\right)
\label{eq:nrho}\\
    N_{\rho y} &= x_0q  \left(\frac{(1+\kappa)^{q-1} - (1-y\kappa)^{q-1}}{(1+y)^2} + \frac{y\kappa(q-1)(1-y\kappa)^{q-2} }{1+y}
    \right) \\
    N_{\rho \rho} &= x_0^2 q(q-1) \frac{y}{1+y}\left((1+\kappa)^{q-2} + y(1-y\kappa)^{q-2}\right) \\
    N_{\rho \rho y} &= x_0^2 q(q-1)\left(
    \frac{(1+\kappa)^{q-2} + y(y+2)(1-y\kappa)^{q-2}}{(1+y)^2} - \frac{(q-2)y^2\kappa (1-y\kappa)^{q-3}}{1+y}
    \right)
\end{align}

\begin{proof}[Proof of \Cref{lem:NNN}]
$N_\rho>0$ is clear from~\eqref{eq:nrho}.
On the other hand,
Let $T(y):=(1+y)^2 N_y$. We have
$T(0)=(1+\kappa)^q-1-q\rho x_0> 0$. At the same
time $T'(y)=q(q-1)\kappa^2(1+y)(1-y\kappa)^{q-2}\ge 0$.
That implies $T(y)> 0$ and $N_y> 0$.
\end{proof}

Using the calculations above, we can see that for fixed $y,x_0>0$
it holds
$N=1+O(\rho)$, $N_y,N_\rho=O(\rho)$ and
$N_{\rho y}=q(q-1)x_0^2\rho+O(\rho^2)$. Consequently,
$\frac{M}{W-M}-L=O(\rho)$, in particular it tends to 0
as $\rho \to 0^+$.
Therefore, in order to prove~\eqref{eq:63}, it is enough 
that $\frac{M}{W-M}-L$ is strictly increasing in~$\rho$. 
Thus, taking derivatives, it suffices to show
\begin{align}
  0 &< \frac{M_\rho W - M W_\rho}{(W-M)^2} - \frac{N_\rho}{N}\;,
  \label{eq:Nybeforefinal}
\end{align}
which is implied by
\begin{align}
    0 &< N(M_\rho W - M W_\rho) - N_\rho(W-M)^2\\
    &= N_y \cdot \left(N_\rho^2(NN_{\rho y}-N_y N_\rho) 
    +N^2 (N_{\rho y}N_{\rho \rho} - N_\rho N_{\rho\rho y})\right).   
\end{align}
Given \Cref{lem:NNN},
it suffices to show the following:
\begin{equation} \label{eqn:Goal}
  \text{Goal:}\qquad
  NN_{\rho y}-N_yN_\rho>0
  \quad\text{ and }\quad
     N_{\rho y}N_{\rho \rho} - N_\rho N_{\rho\rho y}
    \geq 0.
\end{equation}
Note that the first goal $N N_{\rho y}-N_yN_\rho> 0$ implies
$(W-M)^2>0$, so we do not need to worry about division by zero
in~\eqref{eq:Nybeforefinal}.

Let us tidy up a little more, by writing $p \coloneqq q-2 \geq 0$ and dividing out some common factors. Accordingly, let us define
\begin{align}
    n &\coloneqq (1+y) N =(1+\kappa)^{p+2}y + (1-y\kappa)^{p+2} \\
    n_y &\coloneqq (1+y)^2N_y=(1+\kappa)^{p+2} - (1-y\kappa)^{p+1}
    \left(1+(p+2)\kappa+(p+1)y\kappa\right)\\
    n_\rho &\coloneqq \frac{(1+y)}{x_0q }N_\rho =(1+\kappa)^{p+1}y - (1-y\kappa)^{p+1}y\\
    n_{\rho y} &\coloneqq \frac{(1+y)^2}{x_0q}N_{\rho y}= (1+\kappa)^{p+1} - (1-y\kappa)^{p}\left(
    1-(p+2)y\kappa-(p+1)y^2\kappa\right)
     \\
    n_{\rho \rho} &\coloneqq \frac{1+y}{x_0^2q(q-1)}N_{\rho\rho} =(1+\kappa)^{p}y + (1-y\kappa)^{p}y^2 \\
    n_{\rho \rho y} &\coloneqq \frac{(1+y)^2}{x_0^2q(q-1)}=
    (1+\kappa)^{p} 
    +(1-y\kappa)^{p-1}\left(2y+y^2-(p+2)y^2\kappa-(p+1)y^3\kappa\right),
\end{align}
and it is readily checked that 
our goal in~\eqref{eqn:Goal} is implied
by an updated goal
\begin{equation}
n n_{\rho y}-n_y n_\rho>0
\quad\text{ and }\quad
\frac{1}{y}\left(n_{\rho y}n_{\rho\rho}-n_\rho n_{\rho\rho y}\right)\ge 0\;.
\end{equation}
Therefore, to conclude the proof of \Cref{lem:reduction-to-minus-one} it is sufficient to state and prove the two claims below:

\begin{claim}\label{cl:nrhoy-first}
$nn_{\rho y}-n_y n_{\rho}>0$
for $p\ge 0, \kappa>0$ and $0<y<1/\kappa$. 
\end{claim}

\begin{proof}
Let $A\coloneqq(1+\kappa)^{p+1}$ and $B\coloneqq(1-y\kappa)^p$. 
Then,
\begin{align}
&n n_{\rho y}-n_yn_\rho
=\\
&\qquad\left[A(1+\kappa)y+B(1-y\kappa)^2\right]
\cdot\left[
A-B\left(
    1-(p+2)y\kappa-(p+1)y^2\kappa\right)
\right]\\
&\qquad\quad-
\left[A(1+\kappa)-B(1-y\kappa)\left(1+(p+2)\kappa+(p+1)y\kappa\right)\right]\cdot
\left[Ay-B(1-y\kappa)y\right]\;.
\end{align}
Multiplying this out, the coefficient of $A^2$ vanishes, and
the coefficient of $B^2$ is $-(1-y\kappa)^2(1+y)$. A calculation reveals
that the coefficient of $AB$ is equal to
$(1+y)(1+py\kappa+(p+1)y^2\kappa)$. Therefore, substituting back for
$A$ and $B$,
\begin{align}
&nn_{\rho y}-n_yn_\rho
=(1-y\kappa)^p(1+y)\cdot\\
&\qquad\cdot\Big[(1+\kappa)^{p+1}
\left(1+py\kappa+(p+1)y^2\kappa\right)
-(1-y\kappa)^{p+2}
\Big]>0\;.\qedhere
\end{align}
\end{proof}

\begin{claim}\label{cl:nrhoy-second}
$\frac{1}{y}\left(n_{\rho y}n_{\rho\rho}-n_\rho n_{\rho\rho y}\right)>0$
for $p\ge 0, \kappa>0$ and $0<y<1/\kappa$.
\end{claim}
\begin{proof}
The structure of the argument is identical as in the proof
of \Cref{cl:nrhoy-first}. This time, let
$A\coloneqq (1+\kappa)^{p}$ and $B\coloneqq (1-y\kappa)^{p-1}$.
Then,
\begin{align}
&\frac{1}{y}\left(n_{\rho y} n_{\rho\rho}-n_\rho n_{\rho\rho y}\right)
=\\
&\qquad\left[
A(1+\kappa)-B(1-y\kappa)\left(1-(p+2)y\kappa-(p+1)y^2\kappa\right)
\right]
\cdot\left[
A+B(1-y\kappa)y
\right]\\
&\qquad\quad-
\left[
A(1+\kappa)-B(1-y\kappa)^2
\right]\cdot
\left[
A+B\left(2y+y^2-(p+2)y^2\kappa-(p+1)y^3\kappa\right)
\right]\;.
\end{align}
After multiplying out, the coefficient of $A^2$ is zero,
and the coefficient of $B^2$ is $(1+y\kappa)^2y(1+y)$. Finally,
the coefficient of $AB$ is checked to be $-y(1+y)(1-p\kappa-(p+1)y\kappa)$.
Hence,
\begin{align}
&\frac{1}{y}\left(n_{\rho y}n_{\rho\rho}-n_\rho n_{\rho\rho y}\right)
=(1-y\kappa)^{p-1}y(1+y)\cdot
\label{eq:65}\\
&\qquad\cdot\Big[-(1+\kappa)^{p}
\left(1-p\kappa-(p+1)y\kappa\right)
+(1-y\kappa)^{p+1}
\Big]\;.
\label{eq:64}
\end{align}

It remains to show that the term in square brackets in~\eqref{eq:64}
is strictly positive. This follows since it is zero for $\kappa=0$,
and since its derivative with respect to $\kappa$ is
\begin{align}
(1+\kappa)^{p-1}(p+1)\left(
y+p\kappa+(p+1)y\kappa
\right)
-(1-y\kappa)^{p+1}(p+1)y > 0\;.
\end{align}
Therefore, indeed the right hand side of~\eqref{eq:65}--\eqref{eq:64} is 
strictly positive for $\kappa>0$.
\end{proof}

\subsubsection{The function $R_q(\rho,x,1/x)$ is increasing in $x$}
\label{sec:fixed-minus-one}

\begin{proposition}
\label{prop:step-3}
Let 
 $q\ge 2$, $0<\rho<1$ and 
$R(y)\coloneqq R_q(\rho,x,1/x)$.
Then, $R(x)$ is strictly increasing in $x>0$.
\end{proposition}

\begin{proof}
Let $N_\rho(x)\coloneqq N_q(\rho,x,y)$
(see~\eqref{eq:71})
and 
$L_\rho(x)\coloneqq \ln N_\rho(x)$.
Computing the derivative of $R(x)$ with respect to~$x$, and omitting the nonnegative denominator, we get
\begin{align}
    \frac{d}{dx} R(x) \text{ has the same sign as} &\ L_\rho'(x) L_1(x) - L_\rho(x) L_1'(x) 
\end{align}
which is positive (and hence $R(x)$ is increasing, as desired) provided 
\begin{align}
    L'_{\rho}(x) L_1(x) &>  L'_1(x) L_\rho(x) \\
    \iff \quad L'_{\rho}(x) \cdot \frac{L_1(x)}{L'_1(x)} - L_\rho(x) > 0
    \\
    \iff \quad U_\rho(x) \coloneqq L'_{\rho}(x) \cdot (1+x)\ln(1+x) - L_\rho(x) &> 0, 
\end{align}
where we used the specific form of $L_1(x)$.
Since $L_\rho(0) = \ln 1 = 0$,  we get $U_\rho(0) = 0$, and hence it suffices to show $U'_\rho(x)> 0$
for $x>0$.
Now
\begin{align}
    U'_\rho(x) &= L''_\rho(x) \cdot (1+x)\ln(1+x) + L'_\rho(x) \cdot (1 + \ln(1+x)) - L'_\rho(x) \\
    &= \ln(1+x) \cdot (L'_\rho(x) + (1+x) L''_\rho(x)),
\end{align}
and hence it remains to show the following rational inequality:
\begin{equation}
    L'_\rho(x) + (1+x)L''_\rho(x)> 0.
\end{equation}
In turn, this is equivalent to 
\begin{align}
    \frac{N_\rho'(x)}{N_\rho(x)} + (1+x) \frac{N''_\rho(x) N_\rho(x) - N'_\rho(x)^2}{N_\rho(x)^2} &> 0 \\
    \iff \quad 
    V_\rho(x) \coloneqq N_\rho(x)N_\rho'(x) + (1+x)(N''_\rho(x) N_\rho(x) - N'_\rho(x)^2) &> 0 \label{eqn:V}
\end{align}

Let us multiply $V_\rho(x)$ by $(1+x)^2$ (which doesn't affect its sign) and make the substitutions $y\coloneqq 1+\rho x> 1$ and $z\coloneqq 1-\rho \in [0,1]$. (We will also use $x = \frac{y-1}{1-z}$.) 
Thus we are reduced to establishing, for $q \geq 2$, that
\begin{multline}
    0 < U(y,z) \coloneqq (y^q - q(y-z)y^{q-1} - z^q)(q(1-z)y^{q-1}+z^q) \\
    +q(q-1)(y-z)(y^q(1-z)+(y-1)z^q)y^{q-2}.
\end{multline}
In order to do that, we differentiate with respect to $y$. Accordingly,
\begin{align}
\frac{\partial U}{\partial y}
&=q(q-1)y^{q-3}\left[
2y^qz(1-z)+(q-1)y^2z^q-(q-3)yz^{q+1}-qyz^q+(q-2)z^{q+1}
\right]\\
&=:q(q-1)y^{q-3} U_1(y,z)\;,\\
\frac{\partial U_1}{\partial y}
&=2qy^{q-1}z(1-z)+2(q-1)yz^q-(q-3)z^{q+1}-qz^q\\
&> z^q\left( q-2-(q-3)z\right)\ge 0\;,
\end{align}
where in the last step we used $q\ge 2$ and therefore
$q-2-(q-3)z\ge \min(q-2, 1)\ge 0$\;.
As
$U_1(1, z)=(1-z)(2z-z^q)\ge 0$, that implies $U_1(y,z)>0 0$
and $\frac{\partial U}{\partial y}> 0$. Finally,
\begin{align}
U(1,z)&=z\left(q-qz-(2q-1)z^{q-1}+2qz^{q}-z^{2q-1}\right)=:zQ(z)\;.
\end{align}
As $Q'(z)=-q+z^{q-2}(-(q-1)(2q-1)+2q^2z-(2q-1)z^q)$
and $\frac{\mathrm{d}}{\mathrm{d}z}(2q^2z-(2q-1)z^q)\ge 0$,
it follows $Q'(z)\le Q'(1)=0$, note that here we used $q\ge 2$.
Then $Q(z)\ge Q(1)=0$ and $U(1, z)\ge 0$.
Together with $\frac{\partial U}{\partial y}> 0$, which we already
established, that gives $U(y, z)> 0$ for
$y>0$ and $0<z<1$,
which concludes the proof.
\end{proof}
\subsubsection{Proof of \Cref{thm:general-q-rv}}
By \Cref{thm:binary-reduction},
\Cref{lem:reduction-to-minus-one}, and
\Cref{prop:step-3}, the unique maximizer
of $R_{q,\rho}$ over $\mathcal{P}^*_\alpha$
is the random variable achieving the value
$R_q(\rho,\alpha,1/\alpha)$ (see~\eqref{eq:71}).
But this is the random variable with binary support
taking values $\alpha$ and $-1$, that is $X_\alpha$.
\qed

\section{Weight distribution of linear codes}
\label{sec:coding}

In this section we develop an application
of \Cref{thm:tensorization-q-norm} to the theory
of error-correcting codes. 
We will show bounds on a weight distribution of a linear code
in terms of conditional entropies
(of the code or of its dual)
on the erasure channel.
As a consequence, under a mild assumption on minimum distance, if a linear code
over a finite field has vanishing
block decoding error probability on a $k$-ary erasure channel,
it also has vanishing block (MAP) error probability on a wide range
of other communication channels (with some loss in capacity).
The results and proofs follow closely previous
works for the binary case~\cite{Sam19, HSS21}.

We give definitions and background for coding theory as necessary
to state our results. For a more extensive treatment, see,
e.g., \cite[Part IV]{PW25}, \cite{GRS23, RU08}.

\begin{definition}[Communication channel]
A \emph{channel} $W$ with finite input alphabet $\Omega$
and output alphabet $\mathcal{Y}$
(which we will also write as
$W:\Omega\to\mathcal{Y})$
is a collection
of $\mathcal{Y}$-valued\footnote{
One does not lose much by focusing on the case of finite $\mathcal{Y}$,
but we allow $\mathcal{Y}$ to be any measurable space.
} transition probability distributions $W(\cdot|x)$ 
for all $x\in\Omega$.
\end{definition}

\begin{definition}[Erasure channel]
Let $k\ge 2$ and $0 \leq \lambda \leq 1$. 
We let $\kEC(\lambda):\mathbb{Z}_k\to\mathbb{Z}_k\cup\{?\}$
be determined by transition probabilities $W(x|x) = 1 - \lambda$ and $W(?|x) = \lambda$ for every $x\in\mathbb{Z}_k$.
We refer to this channel as  the 
\emph{$k$-ary erasure channel with erasure probability $\lambda$}.
\end{definition}

\begin{definition}[Error-correcting code]
Let $\Omega$ be a finite input alphabet. An (error-correcting) 
\emph{code}
with \emph{block length} $n$ is any nonempty subset
$\cC\subseteq \Omega^n$. If $\Omega=\mathbb{F}_k$ for a finite field $\mathbb{F}_k$ and $\cC$ is a linear subspace of
$\mathbb{F}_k^n$, then we call the code \emph{linear}.

For $x\in\mathbb{F}_k^n$, we let
$\wt(x)\coloneqq \left|\{1\le i\le n: x_i\ne 0\}\right|$.
The \emph{weight distribution} of a linear code $\cC\subseteq \mathbb{F}_k^n$ is the vector
$(a_0,\ldots,a_n)$, where
$a_i\coloneqq |\{x\in\cC: \wt(x)=i\}|$.
The \emph{minimum distance}
of such a code is
$d(\cC)\coloneqq \min\left\{\wt(x-x'): x,x'\in\cC, x\ne x'\right\}$.
\end{definition}

Given a code $\cC\subseteq\Omega^n$ and a channel
$W:\Omega\to\mathcal{Y}$, we consider the following random experiment.
$X\in\cC$ is a random variable which is uniform over $\cC$
and $Y\in\mathcal{Y}^n$ is obtained as follows: Conditioned on $X=x$, for every $1\le i\le n$ independently,
$Y_i$ is sampled from the distribution $W(\cdot|x_i)$.
We are then interested in the \emph{MAP (maximum a posteriori) block
decoding} of $Y$:
\begin{definition}[MAP block decoding]
Let $\Xhat:\mathcal{Y}^n\to \cC$ be a 
(measurable) function satisfying
\begin{align}
\label{eq:17}
\Xhat(y)\in\argmax_{x\in \cC}
\Pr[X=x\;\vert\;Y=y]\;.
\end{align}
for every $y\in\mathcal{Y}^n$.
Then, we define the block decoding error
probability of $\cC$ on channel $W$ as
\begin{align}
\label{eq:18}
\PB(\cC,W)&\coloneqq \Pr[\Xhat(Y)\ne X]\;.
\end{align}
\end{definition}
Note that if the arg max in~\eqref{eq:17}
is not unique, the value of $\Xhat(y)$
can be chosen arbitrarily:
The probability $\PB(\mathcal{C},W)$
does not depend on this tiebreaker choice.
Furthermore, $\Xhat$ minimizes the error probability
$\Pr[f(Y)\ne X]$ among all (measurable) decoders $f:\mathcal{Y}^n\to \cC$.

\begin{definition}
We will be using the Shannon entropy $H(X)$
and its conditional version $H(X|Y)$. We define those
with binary logarithm in the formulas.
Furthermore, for $0\le\gamma\le 1$, let
\begin{align}
h_k(\gamma)&:=-(1-\gamma)\log_k(1-\gamma)-\gamma\log_k\frac{\gamma}{k-1}\;,
\end{align}
with the usual convention $0\cdot \log 0=0$.
\end{definition}

The main result we state in this section is an explicit bound
on the weight distribution of $\cC$
in terms of its conditional entropy
on the erasure channel.
This has been given as \Cref{thm:weight-distribution-intro} in the introduction, but we restate it here
for convenience:

\begin{theorem}\label{decoding_thm}
Let $k$ be a prime power, $n\ge 1$, $\cC\subseteq \mathbb{F}_k^n$ a linear code and $0 \leq \lambda \leq 1$. Let $X$ and $Y$ be the input and output
of transmitting $\cC$ over the channel
$\kEC(\lambda)$ as described above.

Then, the weight distribution of $\cC$
satisfies
\begin{align}
\log_2 \sum_{i=0}^n a_i \cdot \left(\frac{k^{\lambda}-1}{k-1}\right)^i \leq H(X|Y)\;.
\label{eq:25}
\end{align}
\end{theorem}

\Cref{decoding_thm} implies bounds on weight distribution
of linear codes in terms of their performance on the erasure channel.

\begin{corollary}
\label{cor:weight-distribution}
Let $\cC\subseteq \mathbb{F}_k^n$ be a linear code over a finite field
with weight distribution $(a_0,\ldots,a_n)$.
Let $0<\lambda\le 1$ and let $\theta:=\frac{k^\lambda-1}{k-1}$.
\begin{enumerate}
\item
Let $X$ and $Y$ be, respectively, 
the input and output of transmitting $\cC$ through $\kEC(\lambda)$.
Then, for every $0\le i\le n$, it holds $a_i\le\theta^{-i}2^{H(X|Y)}$.
\item Similarly, let $X^\perp$ and $Y^\perp$ be the input
and output of transmitting the dual code $C^\perp$ through
$\kEC(\lambda)$. Let $\gamma:=i/n$ and
\begin{align}
F(k,\gamma,\theta)
&:=\begin{cases}
(1-\theta)^\gamma(1+(k-1)\theta)^{1-\gamma}
&\text{if }\gamma<\frac{k-1}{k}(1-\theta)\;,\\
k^{1-h_k(\gamma)}
&\text{if }\frac{k-1}{k}(1-\theta)\le\gamma\le\frac{k-1}{k}(1+\theta)\;,
\\
(1+\theta)^\gamma(1-(k-1)\theta)^{1-\gamma}
&\text{if }\gamma>\frac{k-1}{k}(1+\theta)\;.
\end{cases}
\end{align}
Then, for every $0\le i\le n$, it holds
\begin{align}
a_i\le \frac{|\cC|}{F(k,\gamma,\theta)^n}\cdot 2^{H(X^\perp|Y^\perp)}\;.
\label{eq:41}
\end{align}
\end{enumerate}
\end{corollary}

Let us make two comments regarding \Cref{cor:weight-distribution}.
First, due to sharp threshold results by Tillich and Z\'{e}mor~\cite{TZ00}, if a linear code
with minimum distance $d$ has a vanishing error probability on the erasure
channel $\kEC(\lambda)$, then for any $\lambda'<\lambda$ it holds 
$H(X|Y)\le n\exp(-\Omega(d))$. Therefore, as soon as $d=\omega(\log n)$,
the $2^{H(X|Y)}$ factor is close to 1. The proof of \Cref{cor:kec-to-ksc} below
contains an example of applying this property.

Second, the bound in~\eqref{eq:41} is nontrivial in the sense
that $F(k,\gamma,\theta)>1$ except for the case
$h_k(\gamma)=1$, which is equivalent to $\gamma=\frac{k-1}{k}$.
Furthermore, in the case where
$F(k,\gamma,\theta)=k^{1-h_k(\gamma)}$,
the bound matches the weight distribution of 
a random code up to lower order factors. Indeed,
a random code of size $|\cC|$ has
in expectation
$|\cC|\cdot \frac{(k-1)^i\binom{n}{i}}{k^n}=
|\cC|\cdot k^{nh_k(\gamma)-n+o(n)}$ codewords of weight $i=\gamma n$.

As a second corollary of \Cref{decoding_thm}, we obtain a bound
on the block error probability of decoding $\cC$ on a variety
of channels. To state those results, we need
a couple more definitions regarding
communication channels.

\begin{definition}  \label{def:bhatt}
Let $X$ be uniform over $\Omega$ and $Y$ be the result of a transmission
of $X$ over a channel $W$. We let
\begin{align*}
\Bh(W)\coloneqq \max_{x\ne x'}\EE \left[\sqrt{\frac{\Pr[X=x'|Y]}{\Pr[X=x|Y]}}\;\Bigg\vert\;X=x\right]\;.
\end{align*}
\end{definition}

\begin{remark}
In case of binary memoryless symmetric (BMS) channels, the parameter $\Bh(W)$
is known as the \emph{Bhattacharyya coefficient} of $W$
(see, e.g., \cite[Chapter 4]{RU08} for background on BMS channels)\footnote{
More generally, for $0<t<1$ one can define 
$Z_t(W)\coloneqq\max_{x\ne x'}
\EE \left[\left(\frac{\Pr[X=x'|Y]}{\Pr[X=x|Y]}\right)^t\;\Bigg\vert\;X=x\right]$.
\Cref{block_err} also holds if $\Bh(W)=Z_{1/2}(W)$ is replaced by $\inf_t Z_t(W)$. We restrict our results to
$t=1/2$ for simplicity and since for sufficiently symmetric
channels, including $\kSC$ and BMS channels, it holds
$\Bh(W)\le Z_{t}(W)$.
}.
\end{remark}

\begin{definition}[Symmetric channel]
Let $k\ge 2$ and $0 \leq \eta\leq 1$.
We define the symmetric channel on $\mathbb{Z}_k$
with error probability $\eta$
(denoted $\kSC(\eta):\mathbb{Z}_k\to\mathbb{Z}_k$)
by transition probabilities
$W(x|x)=1-\eta$ and
$W(x'|x)=\frac{\eta}{k-1}$ for $x'\ne x$.
\end{definition}

\begin{corollary}\label{block_err}
Let $k$ be a prime power, $0<\lambda\le 1$,
and $\mathcal{C} \subseteq \mathbb{F}_k^n$ a linear code with minimum distance $d$. Let $X$ and $Y$,
be, respectively, the input and output
of transmitting $\cC$ through $\kEC(\lambda)$.

Let $W:\mathbb{F}_k\to\mathcal{Y}$ be a communication channel and assume that $c\coloneqq  \Bh(W)\frac{k-1}{k^\lambda -1}$ satisfies $c<1$. Then, it holds
\begin{align}
P_B(\mathcal{C}, W) \leq \frac{c^d}{1-c}\cdot 2^{H(X|Y)}\;.
\end{align}
\end{corollary}

\begin{corollary}
\label{cor:kec-to-ksc}
Let $k$ be a prime power and
$\{\cC_n\}_{n}$ a family of linear codes such
that $\cC_n\subseteq \mathbb{F}_k^n$. Assume that their minimum
distances satisfy $d(\mathcal{C}_n)\ge \omega(\log n)$
and that
$\lim_{n\to\infty} \allowbreak\PB(\mathcal{C}_n,\kEC(\lambda))=0$ for
some $0<\lambda<1$.
Then:
\begin{enumerate}
    \item 
    For any channel $W:\mathbb{F}_k\to\mathcal{Y}$
    that satisfies 
    $\Bh(W)<\frac{k^\lambda-1}{k-1}$ it holds
    \begin{align}
    \lim_{n\to\infty}\PB(\cC_n, W)=0.
    \end{align}
    More precisely, it holds
    $\PB(\cC_n,W)\le\exp(-\Omega(d(\cC_n)))$ for sufficiently large $n$.
    \item In particular, the property stated in the first point holds
    for $\kSC(\eta)$ if
    \begin{align}
    \label{eq:21}
    \Bh(\kSC(\eta))&=
    \frac{k-2}{k-1}\eta+2\sqrt{(1-\eta)\frac{\eta}{k-1}}<\frac{k^{\lambda}-1}{k-1}\;. 
    \end{align}
\end{enumerate}
\end{corollary}

By \Cref{cor:kec-to-ksc}, for any channel that satisfies $\Bh(W)<1$,
there exists some $\lambda<1$ such that if a linear code
(with $\omega(\log n)$ minimum distance)
has vanishing error probability
on $\kEC(\lambda)$, then it also has vanishing error probability on $W$.
It is directly seen from the definition that $\Bh(W)<1$ as long as no two
distributions $W(\cdot|x)$ and $W(\cdot|x')$ are equal for $x\ne x'$.
In particular, $\Bh(\kSC(\eta))<1$
except for the uninformative case $\eta=\frac{k-1}{k}$.

\subsection{Example: MAP decoding on symmetric channels}
\label{sec:coding-example}

Recall that the $k$-ary Shannon capacities
of the erasure and symmetric channels are given by
\begin{align}
C(\kEC(\lambda))&=1-\lambda\;,&
C(\kSC(\eta))&=1-h_k(\eta)\;.
\end{align}

\begin{figure}[htp]
    \centering
    \includegraphics[width=.8\textwidth]{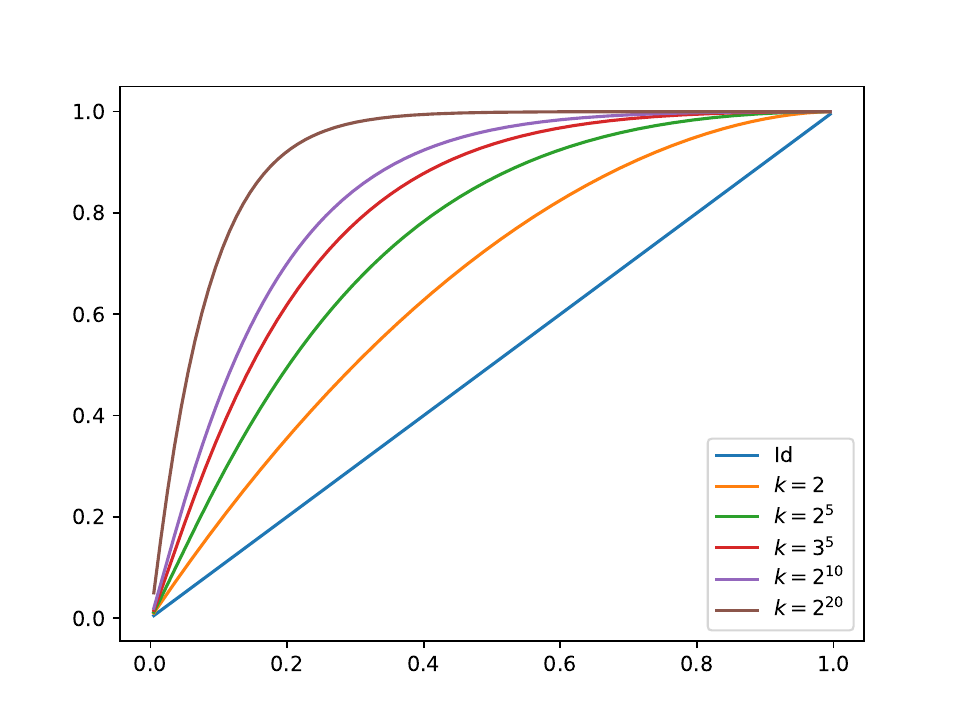}
    \caption{Capacity of $k$-SC$(\eta_\ast)$ versus capacity of $k$-EC$(\lambda)$.}
    \label{fig:cap}
\end{figure}

Given $k$ and $\lambda$,
let $\eta_k(\lambda)$ be the supremum over
$0\le\eta\le\frac{k-1}{k}$ that satisfy \eqref{eq:21}.
In other words,
it is the unique value of $0\le\eta\le\frac{k-1}{k}$ where the
left and right hand sides of~\eqref{eq:21} are equal. Then,
let $g_k(c_e)$ be the capacity of the symmetric channel
$\kSC(\eta^*)$, where $\eta^*=\eta_k(1-c_e)$.
By \Cref{cor:kec-to-ksc}, if $\cC$ is a linear code with vanishing block error
probability on a $k$-ary erasure channel with capacity $c_e$,
then $\cC$ has vanishing block error probability on $k$-ary symmetric channels
with any capacity strictly larger than $g_k(c_e)$.
\Cref{fig:cap} shows the plots of the $g_k$ function for various values of $k$.

Note that for a fixed $0\le \lambda< 1$, it holds
$\lim_{k\to\infty}\frac{k^{\lambda}-1}{k-1}=0$.
It follows $\lim_{k\to\infty} \eta_k(\lambda)=0$
and that $\lim_{k\to\infty} g_k(c_e)=1$ for every $0<c_e\le 1$.
By that measure, the bound from \Cref{cor:kec-to-ksc}
appears to become weaker for fixed $c_e$ as $k$ grows.

On the other hand, clearly we have $g_k(0)=0$
and in \Cref{sec:gk-slope} we 
show that $g'_k(0)=2$ for every
fixed value of~$k$.
Therefore, in the noisy regime of constant $k$
and $\lambda\approx 1$
(and the channel capacity $c_e\approx 0$), we have an approximation 
$g_k(c_e)\approx 2c_e$. In particular, we have

\begin{corollary}\label{cor:asymm}
Let $k$ be a prime power and $\eps>0$. There exists
$\delta_0(k,\eps)>0$ such that for every $\delta\le \delta_0$ the
following holds:

If a family
of linear codes $\cC_n\in\mathbb{F}_k^n$ satisfies both
$d(\cC_n)\ge\omega(\log n)$
and $\lim_{n\to\infty} P_B(\cC_n, \allowbreak\kEC(1-\delta))=0$, then
$P_B(\cC_n, \kSC(\eta))=0$ for $\eta$ satisfying
$C(\kSC(\eta))=2\delta(1+\eps)$.
\end{corollary}

\subsection{Proof of \Cref{decoding_thm}}
For purposes of this proof, we will need
additional tools from discrete Fourier analysis. 
We briefly introduce needed
concepts and refer to, e.g., \cite[Chapter~9]{LN97},
\cite[Chapter~8]{OD14} for more background.
We assume that $k=p^\ell$ is a prime power
and consider the space $L^2(\mathbb{F}_k^n, \mu^n)$, where $\mu$ is uniform
on $\mathbb{F}_k$. Accordingly, we will use shortened notation
$L^2(\mathbb{F}_k^n)$. Furthermore, we will also
work with the space $L^2_\mathbb{C}(\mathbb{F}_k^n)$
of complex-valued function on $\mathbb{F}_k^n$ equipped
with the inner product
\begin{align}
\langle f,g \rangle \coloneqq  \frac{1}{k^n}\underset{x\in \mathbb{F}_k^n}{\sum}f(x)\overline{g(x)} = \underset{x}{\mathbb{E}}\left[f(x)\overline{g(x)}\right]\;.
\end{align}

Given $k$, let us fix a nontrivial
additive homomorphism $\Phi:\mathbb{F}_k\to \mathbb{Z}_p$.
We define the Fourier characters
$\chi_y(x)\coloneqq\exp\left(\frac{2\pi i}{p}\Phi(y\cdot x)\right)$
for $x,y\in\mathbb{F}_k$ and
$\chi_{\alpha}(x)\coloneqq \prod_{j=1}^n\chi_{\alpha_j}(x_j)$
for $\alpha,x\in\mathbb{F}_k^n$.
It is well-known that the functions $\{\chi_{\alpha}\}_{\alpha\in\mathbb{F}_k^n}$ form
an orthonormal basis of $L^2_{\mathbb{C}}(\mathbb{F}_k^n)$.
In particular, for any $f$ it holds $f=\sum_{\alpha\in\mathbb{F}_k^n}\widehat{f}(\alpha)\chi_{\alpha}$,
where $\widehat{f}(\alpha)\coloneqq \langle f,\chi_{\alpha}\rangle$.

\begin{definition}
Given a code $\cC$, we define the function
$f_\cC(x):=\frac{k^n}{|C|}1_\cC(x)$.
For $x,y\in\mathbb{F}_k^n$, let their (pseudo) scalar product
be $\langle x,y\rangle\coloneqq \sum_{i=1}^n x_iy_i$, where the arithmetic is in $\mathbb{F}_k$.
For a code $\cC\subseteq \mathbb{F}_k^n$, its \emph{dual code}
is $\cC^\perp\coloneqq \{x\in\mathbb{F}_k^n: \langle x,y\rangle=0\ \forall y\in\cC\}$.
\end{definition}

(It should  always be possible from context to distinguish between the two scalar products, one on $\mathbb{F}_k^n$ and the other
on the $L^2_\mathbb{C}$ space.)

The following well known claim is proved by an elementary calculation:
\begin{claim}\label{fourier_lem}
Let $\cC \subseteq \mathbb{F}_k^n$ be a linear code.
Then, $\widehat{f_\cC}=1_{\cC^\perp}$.
\end{claim}




For $S\subseteq[n]$, consider the projection operator
$P_S:\mathbb{F}_k^n\to\mathbb{F}_k^S$, where
$P_S(x_1,\ldots,x_n)\coloneqq (x_i)_{i\in S}$. Given a code $\cC\subseteq\mathbb{F}_k^n$,
we let $\cC_S\subseteq\mathbb{F}_k^S$, to be the image
$\cC_S\coloneqq P_S(\cC)$. Clearly, if $\cC$ is a linear code, then so is $\cC_S$
(for any fixed ordering of coordinates in $S$).

Furthermore, recall the notation from \Cref{ex:samorodnitsky}.
For $f\in L^2(\mathbb{F}_k^n)$ and $S\subseteq [n]$, 
the conditional expectation
$\EE(f|S)\in L^2(\mathbb{F}_k^n)$ is defined as
$\EE(f|S)(x)=\frac{1}{k^{n-|S|}}\sum_{y:y_S=x_S}f(y)$.
Recall that 
$\EE(f|S)=T_{\sigma_S} f$, where
$\sigma_{S,i}=1(i\in S)$.

\begin{lemma}\label{norm_lem}
Let $\cC \subseteq \mathbb{F}_k^n$ be a linear code and $f := f_\cC$. Then for $S \subseteq [n]$, we have that
\begin{align}
\EE(f|S)(x) &=\begin{cases}
k^{|S| - \dim \cC_{S}} &\text{if } x_S\in\cC_S, \\
0& \text{otherwise.} 
\end{cases}
\label{eq:23}
\end{align}
In particular, for every $q>1$ it holds 
$\|\EE(f|S)\|_q^{q/(q-1)}=\|\EE(f|S)\|_\infty=k^{|S|-\dim \cC_S}$.
\end{lemma}

\begin{proof}
Indeed, as $P_S$ is a linear operator, and as $\|\EE(f|S)\|_1=\|f\|_1=1$,
it is easy to check~\eqref{eq:23}. In particular $\EE(f|S)(x)=k^{|S|-\dim \cC_S}$
with probability $k^{\dim \cC_S-|S|}$ and $\EE(f|S)(x)=0$ otherwise,
hence $\|\EE(f|S)\|_q^q=k^{(q-1)(|S|-\dim \cC_S)}$.
It is also clear that
$\|\EE(f|S)\|_\infty=k^{|S|-\dim\cC_S}$.
\end{proof}




\begin{lemma}\label{prop:matroid_ox}
Let $\cC \subseteq \mathbb{F}_k^n$ be a linear code of dimension $r$ and 
$S$ a subset of $[n]$. Then $\dim\, (\cC^{\perp})_S = |S| - r + \dim \cC_{S^c}$.
\end{lemma}

\Cref{prop:matroid_ox}
is the formula for the rank of dual matroid applied to the matroid
of columns of a generating matrix of $\cC$. Nevertheless, we give a short
self contained proof.

\begin{proof}
Let $H\in\mathbb{F}_k^{(n-r)\times n}$ be a parity check matrix for $\cC$
and $H_S$ the submatrix of $H$ obtained by keeping only the columns in $S$.
Accordingly, we will think of $H_S$ as a linear mapping from
$\mathbb{F}_k^S$ to $\mathbb{F}_k^{n-r}$. Furthermore, let $P:\cC\to\cC_{S^c}$
be the projection operator mapping codewords of $\cC$ to codewords of 
$\cC_{S^c}$ by deleting the coordinates in $S$. Note that $\dim\ker H_S=\dim\ker P$.
Indeed, $\Phi(x_S)\coloneqq (x_S,0)$ is an isomorphism between $\ker H_S$ and $\ker P$.
If $x_S\in \ker H_S$, then $(x_S,0)\in\mathbb{F}_k^n$ satisfies
$H(x_S,0)=H_S x_S=0$, so $(x_S,0)\in\cC$ and $P(x_S,0)=0$. On the other hand, if
$(x_S,x_{S^c})\in\ker P\subseteq\cC$, then clearly $x_{S^c}=0$ and $x_S\in\ker H_S$.

Applying the rank-nullity theorem to $H_S$, we have $|S|=\dim\ker H_S+\dim\,(\cC^\perp)_S=\dim\ker P+\dim\,(\cC^\perp)_S$.
Doing the same for $P$, we have $r=\dim\ker P+\dim \cC_{S^c}$.
Putting these two together, it follows
$\dim\,(\cC^\perp)_S=|S|-r+\dim \cC_{S^c}$.
\end{proof}

Recall the notation $S\sim\lambda$, which denotes a random set $S\subseteq[n]$
such that each $i$ is included in $S$ independently with probability $\lambda$.

\begin{lemma}\label{matroid_lem}
Let $\cC \subseteq \mathbb{F}_k^n$ be a linear code 
and $X,Y$ be, respectively, input and output of transmitting
$\cC$ over the erasure channel
$\kEC(\lambda)$ for some $0\le\lambda\le 1$. Then,
for every $q> 1$, it holds
\begin{align}
\EE_{S\sim\lambda}\log_k\|\EE(f_{\cC^\perp}|S)\|_q^{q/(q-1)}
&=(\log_k 2)\cdot H(X|Y)\;.
\end{align}
Similarly, it holds $\EE_{S\sim\lambda}
\log_k\|\EE(f_{\cC^\perp}|S)\|_\infty=(\log_k 2)\cdot H(X|Y)$.
\end{lemma}

\begin{proof}
Let $f:=f_{\cC^\perp}$.
By \Cref{prop:matroid_ox}, if $S \subseteq [n]$, then $\dim\,(\cC^\perp)_S = |S| - \dim\cC + \dim \cC_{S^c}$. From this
and~\Cref{norm_lem} it follows
\begin{align}
    \EE_{S\sim\lambda} \log_k \|\EE(f|S)\|_q^{q/(q-1)}
    &=\EE_{S\sim\lambda} \left(|S|-\dim(\cC^\perp)_S\right)
    =\EE_{S\sim\lambda}\left(
    \dim\cC-\dim\cC_{S^c}
    \right)
    \\
    &=\dim\cC-\EE_{S\sim 1-\lambda}\dim\cC_{S}
    =(\log_k 2)\cdot H(X|Y)\;.
    \label{eq:24}
\end{align}
The last line holds due to the structure of MAP block decoding
on an erasure channel: Given $Y=y$, let $S$ bet the set of coordinates
which were not replaced with question marks. Then, the conditional distribution of $X$
is uniform on $P_S^{-1}(y_S)$, which is an affine subspace of
dimension $\dim\ker P_S=\dim\cC-\dim\cC_S$. Therefore,
$H(X|Y=y)=(\log_2 k)\cdot (\dim\cC-\dim \cC_S)$.
Since each coordinate is included in $S$
independently with probability $1-\lambda$, we have
$H(X|Y)=(\log_2 k)\cdot\left(\dim\cC-\EE_{S\sim 1-\lambda}\dim\cC_S\right)$
and~\eqref{eq:24} follows.

The formula for the infinity norm follows by the identical
argument.
\end{proof}

\begin{proof}[Proof of \Cref{decoding_thm}]
    Let
    $f :=f_{\cC^\perp}$ and $\rho \coloneqq  \sqrt{ \frac{k^\lambda -1}{k-1} }$.
     By \Cref{thm:tensorization-q-norm}, and using the reformulation in~\eqref{eq:07},
    \begin{align}
    \log_2 \|T_\rho f\|_2^2 \leq \underset{S \sim \lambda}{\mathbb{E}} \log_2\|\mathbb{E}(f|S)\|_2^2\;.
    \label{eq:26}
    \end{align}
    On the left hand side, using Fourier identities
    $\widehat{T_\rho f}(\alpha) = \rho^{\wt(y)}\widehat{f}(y)$,
    $\|f\|_2^2=\sum_\alpha |\widehat{f}(\alpha)|^2$, and
    \Cref{fourier_lem},
    \begin{align}
    \|T_\rho f\|_2^2&=\sum_{\alpha\in\mathbb{F}_k^n}\rho^{2\wt(\alpha)}
    \left|\widehat{f}(\alpha)\right|^2
    =\sum_{i=0}^n a_i\left(\frac{k^\lambda-1}{k-1}\right)^i\;.
    \label{eq:27}
    \end{align}
    On the right hand side, by \Cref{matroid_lem} applied for $q=2$ it holds
    \begin{IEEEeqnarray}{rCl"s}
       \underset{S \sim \lambda}{\mathbb{E}} \log_k\|\mathbb{E}(f|S)\|_2^2
       & = & (\log_k 2)\cdot H(X|Y)\;.
       \label{eq:44}
    \end{IEEEeqnarray}

    Finally, ~\eqref{eq:25} follows by combining~\eqref{eq:26},
    \eqref{eq:27} and~\eqref{eq:44}.    
\end{proof}

\subsection{Bhattacharyya bound and proofs of the corollaries}

For $\alpha\in\mathbb{R}$, let
$g_\Bh(x):=\alpha^{\wt(x)}$.
The following claim can be checked by an
 elementary calculation:
\begin{claim}
$\widehat{g_\alpha}(y)=\frac{1}{k^n}(1-\alpha)^{\wt(y)}(1+(k-1)\alpha)^{n-\wt(y)}$.
\end{claim}

Together with Parseval identity and \Cref{fourier_lem},
this claim implies the MacWilliams identities for linear codes:
\begin{claim}
\label{cl:macwilliams}
Let $\cC\subseteq\mathbb{F}_k^n$ be a linear code
with weight distribution $(a_0,\ldots,a_n)$
and let $(b_0,\ldots,b_n)$ be the weight distribution
of the dual code $\cC^\perp$. Then, for every
$\alpha\in\mathbb{R}$,
\begin{align}
\sum_{j=0}^n b_j\alpha^j
=\frac{1}{|\cC|}\sum_{j=0}^n a_j(1-\alpha)^j(1+(k-1)\alpha)^{n-j}\;.
\end{align}
\end{claim}


\begin{proof}[Proof of~\Cref{cor:weight-distribution}]
\leavevmode
\begin{enumerate}
\item This bound follows directly from~\Cref{decoding_thm},
as $a_i\theta^i\le\sum_{j=0}^n a_j\theta^j\le 2^{H(X|Y)}$.
\item Let $(b_0,\ldots,b_n)$ be the weight distribution of the dual code
$\cC^\perp$.  
Let $G(\alpha):=(1-\alpha)^{\gamma}(1+(k-1)\alpha)^{1-\gamma}$.
Using \Cref{cl:macwilliams} and \Cref{decoding_thm},
for every $\alpha$ such that
$-\frac{1}{k-1}\le \alpha\le 1$ and $-\theta\le\alpha\le\theta$,
it holds
\begin{align}
a_iG(\alpha)^n
&\le\sum_{j=0}^n
a_j(1-\alpha)^j(1+(k-1)\alpha)^{n-j}\\
&=|\cC|\cdot\sum_{j=0}^n b_j\alpha^j
\le |\cC|\cdot\sum_{j=0}^n b_j\theta^j
\\&\le |\cC|\cdot 2^{H(X^\perp|Y^\perp)}\;.
\label{eq:42}
\end{align}
The bound in~\eqref{eq:41} follows from~\eqref{eq:42} by a case
analysis. If $\gamma<\frac{k-1}{k}(1-\theta)$, then
we take $\alpha:=\theta$ and easily check that $G(\theta)=F(k,\gamma,\theta)$.
If $\frac{k-1}{k}(1-\theta)\le \gamma\le\frac{k-1}{k}(1+\theta)$,
then it follows $-\theta\le 1-\frac{k}{k-1}\gamma\le \theta$
and we can take $\alpha:=1-\frac{k}{k-1}\gamma$.
Then, we check that indeed $G(\theta)=k^{1-h_k(\gamma)}=F(k,\gamma,\theta)$.
Finally, if $\gamma>\frac{k-1}{k}(1+\theta)$, it similarly follows
$-\theta>1-\frac{k}{k-1}\gamma\ge -\frac{1}{k-1}$. Therefore,
we can take $\alpha:=-\theta$.\qedhere
\end{enumerate}
\end{proof}

\begin{remark}
The values of $\alpha$ in the proof of \Cref{cor:weight-distribution}
are chosen with foresight to maximize $G(\alpha)$ over the allowed range
of $\alpha$. In particular, by analyzing its derivative 
it can be checked that $G(\alpha)$
is increasing for $-\frac{1}{k-1}\le\alpha\le 1-\frac{k}{k-1}\gamma$
and decreasing for $1-\frac{k}{k-1}\gamma\le\alpha\le 1$.
\end{remark}

\begin{lemma} \label{z-lemma}
Let $\cC\subseteq \mathbb{F}_k^n$ be a linear code with weight distribution
$(a_0,\ldots,a_n)$. Then, for any channel $W:\mathbb{F}_k\to\mathcal{Y}$
and a block-MAP decoder $\Xhat:\mathcal{Y}^n\to \cC$, for every $x\in \cC$
it holds:
\begin{align}
\Pr[\Xhat(Y)\ne X\;\vert\;X=x]\le\sum_{i=1}^n a_i\Bh(W)^i\;.
\end{align}
\end{lemma}

\begin{proof}
Let $x\in \cC$. By linearity of $\cC$, for every $1\le i\le n$ it holds
\begin{align}
a_i&=\left|\left\{x'\in \cC: \wt(x'-x)=i\right\}\right|\;.
\end{align}
Note that a block-MAP decoder outputs $x$ if $\Pr[X=x|Y]>\Pr[X=x'|Y]$ for
every codeword $x'\ne x$. Therefore, we can apply the union bound over all codewords different
from $x$ and Markov's inequality
(for readability we suppress the conditioning
on $X=x$):
\begin{align}
\Pr[\Xhat(Y)\ne X]
&\le \sum_{\substack{x'\in\cC\\x'\ne x}} \Pr\left[\frac{\Pr[x'|Y]}{\Pr[x|Y]}\ge 1\right]
= \sum_{\substack{x'\in\cC\\x'\ne x}} \Pr\left[\sqrt{\frac{\Pr[x'|Y]}{\Pr[x|Y]}}\ge 1\right]\\
&\le \sum_{\substack{x'\in\cC\\x'\ne x}} \EE\left[\sqrt{\frac{\Pr[x'|Y]}{\Pr[x|Y]}}\right]
=\sum_{\substack{x'\in\cC\\x'\ne x}}\prod_{i=1}^n \EE\left[\sqrt{\frac{\Pr[x_i'|Y_i]}{\Pr[x_i|Y_i]}}\right]
\label{eq:22}
\\
&\le \sum_{\substack{x'\in\cC\\x'\ne x}}\Bh(W)^{\wt(x'-x)}
=\sum_{i=1}^n a_i\Bh(W)^i\;.
\end{align}
The only part that might require further justification
is the equality in~\eqref{eq:22}.
If $\mathcal{Y}$ is finite, then we can
use the fact that $X$ is uniform, Bayes' rule and conditional independence of coordinates
and write
\begin{align}
\sqrt{\frac{\Pr[x'|Y]}{\Pr[x|Y]}}
&=\sqrt{\frac{W(Y|x')}{W(Y|x)}}
=\prod_{i=1}^n\sqrt{\frac{W(Y_i|x'_i)}{W(Y_i|x_i)}}
=\prod_{i=1}^n\sqrt{\frac{\Pr[x'_i|Y_i]}{\Pr[x_i|Y_i]}}\;.
\label{eq:28}
\end{align}
The last equality in~\eqref{eq:28} holds since $\cC$
is a linear code: Either $X_i$ is identically zero, in which case
$x_i=x'_i=0$, or the distribution
of $X_i$ is uniform. In both cases we have $\Pr[X_i=x_i]=\Pr[X_i=x'_i]$
and the  equality follows from the Bayes' rule.
Finally, \eqref{eq:22} holds by taking expectations
 in~\eqref{eq:28} (over $Y$ conditioned on $X=x$) and
 by conditional independence.

\paragraph{General $\mathcal{Y}$.}
To establish~\eqref{eq:22} for a general 
measurable space $\mathcal{Y}$,
we need a slight modification of the argument above. 
Recall that for every $x_i\in\mathbb{F}_k$ we have a probability
distribution $W(\cdot|x_i)$ and consequently for every $x\in\cC$
a product distribution $W(\cdot|x)=\prod_{i=1}^n W(\cdot|x_i)$.
Note that $Y$ is distributed according to the
distribution $\mu\coloneqq\frac{1}{|\cC|}\sum_{x\in\cC} W(\cdot|x)$.
Clearly, for every $x\in\cC$, the distribution $W(\cdot|x)$ is absolutely
continuous with respect to $\mu$.
Therefore, by the Radon--Nikodym theorem, for every
codeword $x$ there exists a function $f_x$
such that $W(A|x)=\int_A f_x\,\mathrm{d}\mu$ for every measurable $A\subseteq\mathcal{Y}^n$.
In particular,
$\frac{1}{|\cC|}\int_Af_x\,\mathrm{d}\mu=\Pr[X=x,Y\in A]$.
Since the conditional probability $\Pr[x|Y]$ also satisfies
$\int_A \Pr[x|Y]\,\mathrm{d}\mu=\Pr[X=x,Y\in A]$ for every $A$,
$\Pr[x|\cdot]=\frac{1}{|\cC|}f_x$ holds $\mu$-a.e., and hence by absolute
continuity also $W(\cdot|x')$-a.e.~for every $x'\in\cC$.

Furthermore, by conditional independence, $f_x(y)=\prod_{i=1}^n f_{x_i}(y_i)$, and by similar considerations
$\Pr[x_i|\cdot]=\frac{1}{k}f_{x_i}$ holds $W(\cdot|x)$-a.e.
(or $f_{x_i}\equiv 1$ in the degenerate case $X_i\equiv 0$).
Going back to~\eqref{eq:22}, it follows by taking
expectations on both sides of the following equality,
which by the foregoing discussion holds
$W(\cdot|x)$-a.e.:
\begin{align}
\sqrt{\frac{\Pr[x'|Y]}{\Pr[x|Y]}}
&=\sqrt{\frac{f_{x'}(Y)}{f_x(Y)}}
=\prod_{i=1}^n\sqrt{\frac{f_{x'_i}(Y_i)}{f_{x_i}(Y_i)}}
=\prod_{i=1}^n\sqrt{\frac{\Pr[x'_i|Y_i]}{\Pr[x_i|Y_i]}}\;.
\qedhere
\end{align}
\end{proof}

\begin{proof}[Proof of \Cref{block_err}]
Let $(a_0, a_1, \ldots, a_n)$ denote the weight distribution of $\mathcal{C}$ and let $h:=H(X|Y)$. Let $\theta = \frac{k^\lambda -1}{k-1} $. By Theorem \ref{decoding_thm}, we have that for all $i$, $\log_2 (a_i \theta^i) \leq h$. So, $a_i \leq 2^h (1/\theta)^i$. By Lemma \ref{z-lemma} and
applying the assumption on the minimum distance,
\begin{align}
P_B(\mathcal{C}, W)& \leq \overset{n}{\underset{i=d}{\sum}} a_i \Bh(W)^i\\
& \leq 2^h \overset{n}{\underset{i=d}{\sum}}\left(\frac{\Bh(W)}{\theta}\right)^i\\
& = 2^h \overset{n}{\underset{i=d}{\sum}} c^i \leq 2^h \frac{c^d}{1-c}\;.\qedhere
\end{align}
\end{proof}

\begin{proof}[Proof of \Cref{cor:kec-to-ksc}]
We justify each of the two statements in turn.
\begin{enumerate}
\item
Recall the alphabet of the erasure channel 
$\mathcal{Y}\coloneqq \mathbb{F}_k\cup\{?\}$. Given a linear code $\mathcal{C}$
and $y\in\mathcal{Y}$,
let $\Cdec(y)\coloneqq \{x\in\mathcal{C}: x_i=\ ?\text{ or }x_i=y_i\ \forall 1\le i\le n\}$.
Let $X$ and $Y$ be input and output when transmitting $\cC$ over an
erasure channel $\kEC(\lambda)$. Then, let us define the
``probability of ambiguous decoding''
\begin{align}
\Pamb(\cC,\lambda)&\coloneqq 
\Pr[|\Cdec(Y)|>1]\;.
\end{align}
Recall that the MAP block decoder $\Xhat$ for $\cC$ on an erasure channel
has a special structure: $\Xhat$ is a MAP block decoder if and only if
$\Xhat(y)\in\Cdec(y)$ for every $y$ that occurs with nonzero probability.
As, conditioned on $Y=y$, the distribution of $X$ is uniform on
$\Cdec(y)$, we also have
\begin{align}
\Pr[\Xhat(Y)= X\;\vert\;Y=y]
&=\frac{1}{|\Cdec(y)|}\;.
\end{align}
In particular, if $|\Cdec(y)|=1$, then there is no decoding error,
and consequently $\PB(\cC,\kEC(\lambda))\le \Pamb(\cC,\lambda)$.
On the other hand, if $|\Cdec(y)|>1$, then $|\Cdec(y)|\ge k$ since $\Cdec(y)$ is an affine
subspace of $\cC$. As a result,
\begin{align}
\PB(\cC,\kEC(\lambda))
&=
\EE\left[1-\frac{1}{|\Cdec(Y)|}\right]
\ge\frac{k-1}{k}\Pamb(\cC,\lambda)\;.
\end{align}
It follows from this discussion that
$\lim_{n\to\infty}\PB(\cC_n,\kEC(\lambda))=0$
if and only if $\lim_{n\to\infty}\Pamb(\cC_n,\lambda)=0$. 

Given a channel $W$, choose $\lambda'<\lambda$ that satisfies
$c'\coloneqq \Bh(W)\frac{k-1}{k^{\lambda'}-1}<1$. 
Since $\lim_{n\to\infty}\Pamb(\cC_n,\lambda)=0$, certainly
we have $\Pamb(\cC_n,\lambda)\le 1/2$ for large enough $n$.
Applying \cite[Theorem 5.2]{TZ00}, we then have
\begin{align}
\Pamb(\cC_n,\lambda')
&\le\Phi\left(\sqrt{2d}\left(\sqrt{\ln1/\lambda}-\sqrt{\ln 1/\lambda'}\right)\right)
\le\exp(-\Omega(d_n))\;,
\label{eq:19}
\end{align}
where $\Phi$ is the Gaussian CDF, $d_n:=d(\cC_n)$, and the last step follows from
$\Phi(-a)\le\exp(-a^2/2)$ for $a\ge 0$.
\begin{align}
\Pamb(\cC_n,\lambda')\le\exp(-\Omega(d_n))\;,
\end{align}
As already mentioned, if $|\Cdec(y)|=1$, then the erasure channel 
decoding is unique and $H(X|Y=y)=0$. In any other case it certainly
holds $H(X|Y=y)\le n\cdot \log_2 k$. Accordingly, \eqref{eq:19}
implies
\begin{align}
\label{eq:20}
H(X|Y)\le n\cdot \log_2k \cdot \exp(-\Omega(d_n))\;,
\end{align}
where we suppressed $n$ from the notation on the left-hand side.
Due to the assumption $d_n=\omega(\log n)$, it follows
$\lim_{n\to\infty}H(X|Y)=0$, and in particular
$2^{H(X|Y)}\le 2$ for sufficiently large $n$. It remains to
apply \Cref{block_err} to obtain
\begin{align}
\PB(\cC_n,W)\le \frac{2}{1-c'}(c')^{d_n}\;,
\end{align}
which concludes the proof.
\item
The second point is straightforward after checking by direct computation
that
\begin{align}
\Bh(\kSC(\eta))
&=
\frac{k-2}{k-1}\eta+2\sqrt{(1-\eta)\frac{\eta}{k-1}}\;.\qedhere
\end{align}



\end{enumerate}
\end{proof}

\section{Probability of Undetected Error on $k$-Symmetric Channels}
\label{sec:pue}

For a code $\cC \subseteq \mathbb{Z}_k^n$ and $x\in\cC$,
the probability
of undetected error when $x$ is transmitted over a symmetric channel $\kSC(\eta)$ is
the probability that the received message, though different from the transmitted codeword, is also an element of $\cC$. 
If $\cC\subseteq\mathbb{F}_k^n$ is linear with weight distribution
$(a_i)_{i=0,\ldots,n}$, this probability does not depend on $x$ and is
equal to
\begin{align}
\Pue(\cC, \eta) := \sum_{i=1}^n a_i \left(\frac{\eta}{k-1}\right)^i (1-\eta)^{n-i}\;.
\end{align}

For some background on this problem, see, e.g., \cite{KK95}.
For our purposes, let us note only a few facts.
Since in a random linear code, each nonzero element is distributed uniformly
over $\mathbb{F}_k^n\setminus\{0\}$, the expected undetected error probability
of a random linear code is close to (and upper bounded by)
$\frac{|\cC|}{k^n}$, for every value of $\eta>0$. Furthermore,
for codes with sufficiently high rates,
this is essentially the best possible
(proofs are included in \Cref{sec:pue-proofs}):

\begin{lemma}
\label{cl:pue-lower-bound}
For a linear code $\cC$ and $0\le\eta\le\frac{k-1}{k}$, it holds
$\Pue(\cC,\eta)\ge \frac{|\cC|}{k^n}- (1-\eta)^n$.
\end{lemma}

We show that if the dual code has small conditional entropy
on an erasure channel, then the undetected error probability
matches that of the random code for a range of error probabilities:
\begin{theorem}
\label{thm:pue-2-norm}
Let $\cC \subseteq \mathbb{F}_k^n$ be a linear code.
Let $0\le\lambda\le 1$ and  $H(X^\perp|Y^\perp)$ the Shannon entropy of receiving $Y^\perp$ when $X^\perp \in \mathcal{C}^\perp$ is transmitted through the erasure channel $\kEC(\lambda)$. Then,
\begin{align}
\Pue(\mathcal{C}, \eta ) \leq \frac{|\mathcal{C}|}{k^n} \cdot 2^{H(X^\perp|Y^\perp)}
\label{eq:53}
\end{align}
for $1-k^{\lambda-1} \leq \eta\le \min\left(1,1-\frac{2}{k}+k^{\lambda-1}\right)$.
\end{theorem}

\Cref{thm:pue-2-norm} generalizes the result obtained by 
Samorodnitsky in the case of binary linear codes~\cite{Sam22}. While his proof
used the inequality for the infinity norm\footnote{Applying the fact that
for $\rho:=1-\frac{k}{k-1}\eta$ it holds
$\frac{|\cC|}{k^n}T_\rho f_\cC(0)=(1-\eta)^n+\Pue(\cC,\eta)$.}, 
we give a different proof using
\Cref{thm:tensorization-q-norm} for $q=2$.
It can be checked that for $\eta\le\frac{k-1}{k}$ our proof gives the same dependence
of $\eta$ and $\lambda$ as the $q=\infty$ proof.
In the (academic) case of $\eta>\frac{k-1}{k}$
using $q=2$ results in a better bound than applying
\Cref{thm:tensorization-inf-norm} directly.

Furthermore, using Fourier dualities, we can bound the undetected error probability directly in terms of the conditional entropy of $\cC$:
\begin{corollary}
\label{cor:pue-primal}
For $\cC\subseteq\mathbb{F}_k^n$ a linear code,
$0\le\lambda\le 1$, and $X,Y$ input and output of transmitting
$\cC$ over the erasure channel $\kEC(\lambda)$, it holds
\begin{align}
\Pue(\cC,\eta)\le\frac{|\cC|}{k^n}+(1-\eta)^n\cdot 2^{H(X|Y)}\;,
\label{eq:56}
\end{align}
for $0\le \eta\le 1-k^{-\lambda}$.
\end{corollary}

\begin{remark}
By a standard calculation using chain rule, it always holds
$H(X|Y)\ge (\log_2 k)\cdot\left(\dim\cC-(1-\lambda)n\right)$.
Therefore, the second term on the right hand side
of~\eqref{eq:56} satisfies
\begin{align}
(1-\eta)^n\cdot 2^{H(X|Y)}
\ge\frac{|\cC|}{k^n}\cdot(1-\eta)^nk^{\lambda n}
\ge\frac{|\cC|}{k^n}\;,
\end{align}
and the bound in~\eqref{eq:56} matches
the random code bound $\frac{|\cC|}{k^n}$ up to $\exp(o(n))$
factor only if $\eta=1-k^{-\lambda}$
and $H(X|Y)\le(\log_2 k)\cdot\left(\dim\cC-(1-\lambda)n\right)+o(n)$.
\end{remark}

\subsection{Proofs}
\label{sec:pue-proofs}

\begin{lemma}\label{cl:pue-dual}
Let $\cC\subseteq\mathbb{F}_k^n$ be a linear code,
$0\le\eta\le 1$,
and $(b_i)_{i=0,\ldots,n}$ the weight distribution
of the dual code $\cC^\perp$. 
Let $\alpha:=1-\frac{k}{k-1}\eta$.
Then,
\begin{align}
\Pue(\cC,\eta)&=\frac{|\cC|}{k^n}\left(\sum_{i=0}^n b_i\alpha^i\right)
-(1-\eta)^n\;.
\label{eq:57}
\end{align}
In particular, if $0\le\eta\le\frac{k-1}{k}$, then
\begin{align}
\Pue(\cC,\eta)=\frac{|\cC|}{k^n}
+|\cC|(1-\eta)^n\Pue\left(\cC^\perp,\frac{k-1}{k}\cdot
\frac{1-\frac{k}{k-1}\eta}{1-\eta}\right)-(1-\eta)^n\;.
\label{eq:58}
\end{align}
\end{lemma}

\begin{proof}
Let $(a_0,\ldots,a_n)$ be the weight distribution of $\cC$.
\eqref{eq:57} is an application of \Cref{cl:macwilliams}:
\begin{align}
\Pue(\cC,\eta)
&=
\left(\sum_{i=0}^n a_i\left(\frac{\eta}{k-1}\right)^i(1-\eta)^{n-i}\right)
-(1-\eta)^n\\
&=\frac{1}{k^n}\left(\sum_{i=0}^n a_i(1-\alpha)^i(1+(k-1)\alpha)^{n-i}\right)
-(1-\eta)^n\\
&=\frac{|\cC|}{k^n}\left(\sum_{i=0}^n b_i\alpha^i\right)-(1-\eta)^n
\;.
\label{eq:59}
\end{align}
Let $0\le\eta\le\frac{k-1}{k}$ and
$\eta':=\frac{k-1}{k}\cdot\frac{\alpha}{1-\eta}$.
To get~\eqref{eq:58} we continue from~\eqref{eq:59}: 
\begin{align}
\Pue(\cC,\eta)
&=\frac{|\cC|}{k^n}\left(\sum_{i=0}^n b_i\alpha^i\right)-(1-\eta)^n\\
&=\frac{|\cC|}{k^n}+\frac{|\cC|}{k^n(1-\eta')^n}
\left(\sum_{i=1}^n b_i\left(\frac{\eta'}{k-1}\right)^i(1-\eta')^{n-i}\right)-(1-\eta)^n\\
&=\frac{|\cC|}{k^n}+|\cC|(1-\eta)^n \Pue(\cC^\perp,\eta')-(1-\eta)^n\;.
\qedhere
\end{align}
\end{proof}

\begin{proof}[Proof of \Cref{cl:pue-lower-bound}]
Immediately from \Cref{cl:pue-dual}.
\end{proof}

\begin{proof}[Proof of~\Cref{thm:pue-2-norm}]
Let $f:=f_\cC$ and $\alpha:=1-\frac{k}{k-1}\eta$,
$\lambda':=\log_k(1+(k-1)|\alpha|)$.
Check that $\lambda'\le\lambda$.
Combining~\eqref{eq:57}, Parseval, \Cref{fourier_lem}, \Cref{thm:tensorization-q-norm} for $q=2$,
and \Cref{matroid_lem},
\begin{align}
\Pue(\cC,\eta)
&\le \frac{|\cC|}{k^n}\sum_{i=0}^n b_i\alpha^i
\le \frac{|\cC|}{k^n}\sum_{i=0}^n b_i|\alpha|^i
=\frac{|\cC|}{k^n}\left\|T_{\sqrt{|\alpha|}}f\right\|_2^2
\\&\le
\frac{|\cC|}{k^n}\exp_k\left(
\EE_{S\sim\lambda'}\log_k\|\EE(f|S)\|_2^2
\right)
\\&\le
\frac{|\cC|}{k^n}\exp_k\left(
\EE_{S\sim\lambda}\log_k\|\EE(f|S)\|_2^2
\right)
=\frac{|\cC|}{k^n}\exp_2(H(X^\perp|Y^\perp))\;.\qedhere
\end{align}
\end{proof}

\begin{proof}[Proof of \Cref{cor:pue-primal}]
Let $\eta':=\frac{k-1}{k}\cdot\frac{1-\frac{k}{k-1}\eta}{1-\eta}$.
First, check that $\eta\le 1-k^{-\lambda}$
implies $1-k^{\lambda-1}\le\eta'$. Therefore, we can
combine~\eqref{eq:58} and \Cref{thm:pue-2-norm}:
\begin{align}
\Pue(\cC,\eta)
&\le\frac{|\cC|}{k^n}+|\cC|(1-\eta)^n\Pue(\cC^\perp,\eta')\\
&\le\frac{|\cC|}{k^n}+\frac{|\cC|\cdot|\cC^\perp|}{k^n}
(1-\eta)^n 2^{H(X|Y)}
=\frac{|\cC|}{k^n}+(1-\eta)^n\cdot 2^{H(X|Y)}\;.\qedhere
\end{align}
\end{proof}

\bibliographystyle{alpha}
\bibliography{biblio}

\appendix

\section{Proof of \Cref{thm:minkowski}}
\label{app:minkowski-proof}
If $|f(x)|\le |g(x)|$ for every $x$, then
$\|f\|_{x:p}\le \|g\|_{x:p}$ for every $p\in\mathbb{R}\cup\{\infty\}$.
Therefore, 
we only need to prove a two-variable version
\begin{align}
\|f\|_{x:p,y:q}\le\|f\|_{y:q,x:p}\;.
\label{eq:05}
\end{align}
Let us leave the border cases for later and assume $p,q\notin\{0,\infty\}$ and also $f(x,y)\ne 0$ for every
$x,y$.
Define $r\coloneqq q/p$,
and consider both sides of~\eqref{eq:05} raised to the power~$p$; we obtain
\begin{align}
L&\coloneqq\|f\|_{x:p,y:q}^p=\left\|\EE_x|f(x,y)|^p\right\|_{y:r}\;,\\
R&\coloneqq\|f\|_{y:q,x:p}^p=\EE_x\Big\||f(x,y)|^p\Big\|_{y:r}\;.
\end{align}
For $p>0$, we have $r\ge 1$ and $L\le R$ follows by the triangle inequality.
For $p<0$, it can be checked that $r\le 1$ and $L\ge R$ (which is equivalent
to~\eqref{eq:05}) follows by the reverse triangle inequality.

It remains to handle the border cases. Let us keep the assumption
$f(x,y)\ne 0$ for every $x,y$, but allow $p,q\in\{0,\infty\}$. In that case
we use the fact that $\lim_{n\to\infty}\|f\|_{p_n}=\|f\|_p$ if $p_n\to p\in\{0,\infty\}$.
For example, if $p=0<q<\infty$, take any sequence $0< p_n<q$ such that
$\lim_{n\to\infty}p_n=0$. By the first part of this proof, it holds
$\|f\|_{x:p_n,y:q}\le \|f\|_{y:q,x:p_n}$ for every $n$.
Using $\lim_{n\to\infty}\|f\|_{p_n}=\|f\|_0$ and the continuity of
the $q$-norm, we obtain
$\|f\|_{x:0,y:q}\le\|f\|_{y:q,x:0}$.
Other cases of $p$ and $q$ are handled similarly.

Finally, consider $f(x,y)$ that can take zero values.
Let $f_\eps(x,y)$ be equal to $f$ if $f(x,y)\ne 0$ and equal to $\eps$ if
$f(x,y)=0$. It can be checked that
$\lim_{\eps\to 0}\|f_\eps\|_{x:p,y:q}=\|f\|_{x:p,y:q}$ for each
$p,q\in\mathbb{R}\cup\{\infty\}$. Accordingly, we get~\eqref{eq:05}
by a limiting argument.\qed

\section{$g_k'(0)=2$}
\label{sec:gk-slope}

\begin{proposition}
\label{prop:gk-slope}
Let $k\geq 2$. The slope of $g_k$ at zero is 2.
\end{proposition}

\begin{proof}
Since $g_k(0)=0$, showing $g_k'(0)=2$ is equivalent
to $g(\eps)=2\eps+o(\eps)$.
Accordingly,
suppose $C(\kEC(\lambda)) = \epsilon$, so that $\lambda = 1 - \epsilon$. Then $C(\kSC(\eta^*)) = 1 -h_k(\eta^*)$, where $\eta^*$ is the solution 
over $0\le\eta\le \frac{k-1}{k}$ to 
\begin{align}\label{eq:cap}
\frac{k^{1-\epsilon} -1}{k-1} = \left(\frac{k-2}{k-1}\right)\cdot \eta + 2\sqrt{(1-\eta)\cdot \frac{\eta}{k-1}}  
\end{align}

As $k^{1-\epsilon} = k \cdot k^{-\epsilon} = k \cdot e^{-\epsilon \cdot \ln k} = k(1 -\epsilon \ln k + O(\epsilon^2))$, the LHS of \Cref{eq:cap} is equal to 
$1 - \epsilon \frac{k\ln k}{k-1} + O(\epsilon^2)$.

To evaluate the RHS, suppose that $\eta = \frac{k-1}{k} - \delta$. The RHS of \Cref{eq:cap} becomes
\begin{align}
\text{RHS}&=
 \frac{k-2}{k-1}\left( \frac{k-1}{k} - \delta \right) + 2 \sqrt{\left(\frac{1}{k} + \delta \right)\left(\frac{1}{k} - \frac{\delta}{k-1}\right)} \\
 &= \frac{k-2}{k-1}\left( \frac{k-1}{k} - \delta \right) + \frac{2}{k} \sqrt{(1 + k\delta)\left(1 - \frac{k}{k-1}\delta\right)}\\ 
&  =\frac{k-2}{k} - \frac{k-2}{k-1}\delta + \frac{2}{k} \sqrt{ 1 + \frac{k(k-2)}{k-1}\delta - \frac{k^2}{k-1}\delta^2}\\ 
 &=\frac{k-2}{k} - \frac{k-2}{k-1}\delta + \frac{2}{k}\left(1 + \frac{k(k-2)}{2(k-1)}\delta - \frac{k^2}{2(k-1)}\delta^2 - \frac{1}{8}\frac{k^2(k-2)^2}{(k-1)^2}\delta^2 + O(\delta^3) \right)\\
 &= 1 - \frac{k^3}{4(k-1)^2}\delta^2 + O(\delta^3).
\end{align}
Therefore, given $\eps>0$, the unique solution $\eta^*=\frac{k-1}{k}-\delta$
satisfies
\begin{align}
    1 -  \frac{k\ln k}{k-1}\epsilon + O(\epsilon^2) = 1 - \frac{k^3}{4(k-1)^2}\delta^2 + O(\delta^3)\;,
\end{align}
and therefore must satisfy
$\delta^2 = \frac{4(k-1)\ln k}{k^2}\epsilon + o(\epsilon)$.
The formula for capacity of $\kSC(\eta)$ becomes:

\begin{align}
g_k(\epsilon) &=
1-h_k\left(\frac{k-1}{k}-\delta\right)\\
&= 1 + \left(\frac{1}{k} + \delta\right) \log_k\left(\frac{1}{k} + \delta\right) + \left(\frac{k-1}{k} - \delta\right) \log_k\left( \frac{1}{k} - \frac{\delta}{k-1}\right)\\
&=\frac{1}{\ln k}\left(  \left(\frac{1}{k} + \delta\right) \ln (1 + k\delta) + \left(\frac{k-1}{k} - \delta\right)\ln\left(1 - \frac{k}{k-1}\delta\right)  \right)\\
&=\frac{1}{\ln k}\Bigg( \left(\frac{1}{k} + \delta\right)\left( k\delta - \frac{k^2\delta^2}{2} + O(\delta^3) \right)\\ 
&\qquad\qquad+ 
\left(\frac{k-1}{k} - \delta\right) \left( -\frac{k}{k-1}\delta - \frac{k^2}{2(k-1)^2}\delta^2 + O(\delta^3)\right)   \Bigg)\\
&=\frac{1}{\ln k}\left( \frac{k}{2}\delta^2 + \frac{k}{2(k-1)}\delta^2 + O(\delta^3)  \right)\\
&= \frac{k^2}{2(k-1)\ln k}\delta^2 + O(\delta^3)\\
&= \frac{k^2}{2(k-1)\ln k} \cdot \frac{4(k-1)\ln k}{k^2}\epsilon + o(\epsilon)\\
&= 2\epsilon + o(\epsilon^2)\;.
\end{align}
As mentioned, that implies $g'(0)=2$.
\end{proof}

\end{document}